%% file: main.tex
\documentclass[acmsmall]{acmart}

\setcopyright{cc}
\setcctype{by}
\acmJournal{PACMHCI}
\acmYear{2025} \acmVolume{9} \acmNumber{7} \acmArticle{CSCW396} \acmMonth{11} \acmPrice{}\acmDOI{10.1145/3757577}

\AtBeginDocument{%
  }
    
\usepackage{multirow}
\usepackage{xcolor}
\newcommand{\rzre}[1]{\textcolor{black}{#1}}
\newcommand{\rblue}[1]{\textcolor{black}{#1}}
\newcommand{\mt}[1]{\textcolor{black}{#1}}
\usepackage[normalem]{ulem}

\title[Designing Effective AI Explanations for Misinformation Detection]{Designing Effective AI Explanations for Misinformation Detection: A Comparative Study of Content, Social, and Combined Explanations }



\received{October 2024}
\received[revised]{April 2025}
\received[accepted]{August 2025}

\author{Yeaeun Gong}
\affiliation{%
  \institution{University of Illinois Urbana-Champaign}
  \country{USA}}
  \email{yegong2@illinois.edu}

\author{Yifan Liu}
\affiliation{%
  \institution{University of Illinois Urbana-Champaign}
  \country{USA}}
  \email{yifan40@illinois.edu}

\author{Lanyu Shang}
\affiliation{%
  \institution{Loyola Marymount University}
  \country{USA}}
\email{lanyu.shang@lmu.edu}

\author{Na Wei}
\affiliation{%
  \institution{University of Illinois Urbana-Champaign}
  \country{USA}}
\email{nawei2@illinois.edu}

\author{Dong Wang}
\affiliation{%
  \institution{University of Illinois Urbana-Champaign}
  \country{USA}}
\email{dwang24@illinois.edu}

\begin{document}
\input{0abstract.tex}

\maketitle

\graphicspath{{pic/}}
\input{1introduction.tex}
\input{2related.tex}

\input{3gpt}
\input{3study1}
\input{4study2}
\input{5discussion}
\input{6conclusion}

\section{Acknowledgments}
This research is supported in part by the National Science Foundation under Grant No. CNS-2427070,  IIS-2331069,  IIS-2202481, IIS-2130263, CNS-2131622. The views and conclusions contained in this document are those of the authors and should not be interpreted as representing the official policies, either expressed or implied, of the U.S. Government. The U.S. Government is authorized to reproduce and distribute reprints for Government purposes notwithstanding any copyright notation here on.

\bibliographystyle{ACM-Reference-Format}
\bibliography{ref}

\newpage
\appendix
\input{7appendix}

\end{document}

%% file: 0abstract.tex
\begin{abstract}
In this paper, we study the problem of AI explanation of misinformation, where the goal is to identify explanation designs that help improve users’ misinformation detection abilities and their overall user experiences. Our work is motivated by the limitation of current Explainable AI (XAI) approaches, which predominantly focus on content explanations that elucidate the linguistic features and sentence structures of the misinformation. To address this limitation, we explore various explanations beyond content explanation, such as ``social explanation'' that considers the broader social context surrounding misinformation, as well as a ``combined explanation'' where both the content and social explanations are presented in scenarios that are either aligned or misaligned with each other. To evaluate the comparative effectiveness of these AI explanations, we conduct two online crowdsourcing experiments in COVID-19 (Study 1 on Prolific) and Politics domains (Study 2 on MTurk). Our results show that AI explanations are generally effective in aiding users to detect misinformation, with effectiveness significantly influenced by the alignment between content and social explanations. We also find that the order in which explanation types are presented—specifically, whether a content or social explanation comes first—can influence detection accuracy, with differences found between the COVID-19 and Political domains. \rzre{This work contributes towards more effective design of AI explanations, fostering a deeper understanding of how different explanation types and their combinations influence misinformation detection.}
\end{abstract}

%% file: 1introduction.tex
\section{Introduction}
\label{sec:intro}

The spread of online misinformation on social media has been a severe societal issue, with about 74\% of U.S. adults viewing its proliferation as a major threat.\footnote{\url{https://apnews.com/article/religion-crime-social-media-race-and-ethnicity-05889f1f4076709c47fc9a18dbee818a}} This concern has led to a surge in research and public interest, leading to significant progress in the development of artificial intelligence (AI) technologies that aim at detecting and combating misinformation. However, despite the advancements in misinformation detection models, translating these detection results to end users in a non-technical, user-friendly way remains a challenge \cite{liao2021human}. Studies have shown that explanations from common technical methods are often complex and cognitively demanding, thus limiting their usability in real-world settings \cite{jacobs2021designing,springer2019progressive}. For example, complex explanations can lead users to more likely experience cognitive overload, and therefore lack the motivation to fully understand AI systems and critically analyze the explanations provided \cite{wang2021explanations}. Given that one of the goals of explainable AI (XAI) is to enhance human understanding of AI models and empower users for informed decision-making \cite{liao2021human}, it is crucial to develop effective and user-friendly explanation designs. In this paper, we study the problem of \textit{AI explanation of misinformation}, where the goal is to identify and investigate explanation that improves users’ misinformation detection abilities and their overall user experience.

Recent efforts have been made to develop user-friendly explanations for misinformation detection systems. Examples of these explanations include human-understandable explanations through visualizations \cite{yang2019xfake} and interfaces that allow users to explore and interact with the decision-making processes of misinformation detectors \cite{mohseni2021machine}. More recently, there has been an increasing emphasis on natural language explanations, aimed at making these explanations more compatible with how humans naturally process information and articulate their reasoning 
\cite{kou2022crowd}. However, while these explanations have shown effectiveness in enhancing user understanding of AI and misinformation detection accuracy, there are still two critical research gaps that need to be addressed.  

First, existing explanations primarily focus on \textbf{\textit{content explanations}} that detail the linguistic features and sentence structures of the misinformation \cite{yang2019xfake, mohseni2021machine}. Yet, there is a lack of work on \textbf{\textit{social explanations}} that explain the vital social contexts behind misinformation. These social contexts include cues such as user attributes (e.g., age, gender) \cite{joy2021you}, user engagement metrics (e.g., number of likes and shares) \cite{guo2018rumor}, information diffusion patterns (e.g., wider and rapid propagation with repeated surges) \cite{liu2018early}, and user comments that flag misinformation \cite{shang2022duo}. Given that misinformation detection models incorporate both content and socio-contextual cues, there's a clear gap between research on detecting misinformation and explaining the detection process. A few recent studies in computer-supported cooperative work (CSCW) and human computer interaction (HCI) communities have examined whether providing social contexts of misinformation, such as tweet trajectories and user attributes, aids individuals in identifying misinformation \cite{zade2023tweet,waltenberger2023reddit, im2020synthesized}.
While these studies offer insights into the benefits of social explanations in enhancing users' misinformation detection abilities by evaluating the credibility of sources, they mainly used qualitative methods 
and did not quantitatively assess the effectiveness of the social explanations. Thus, a more comprehensive study that explores the effect of various explanations for misinformation, encompassing \textit{both} content and social explanations, is needed.

Second, the impact of the \textbf{\textit{alignment}} between content and social explanations on users’ misinformation detection abilities remains unexplored. 
This gap opens new avenues for XAI research, especially in scenarios where explanations conflict. Consider the statement, ``The COVID-19 pandemic has deeply shaken our world, stirring widespread fear and sorrow.'' A content explanation might label this statement as misleading, highlighting the use of evocative language (e.g., `deeply shaken', `stirring widespread fear and sorrow') that are commonly employed in misinformation to manipulate public emotions and skew perception \cite{carrasco2022fingerprints}. On the other hand, a social explanation might label it as true, noting its widespread sharing on trusted health forums. Since misinformation is influenced not only by content-related, but also by socio-contextual factors \cite{ecker2022psychological}, many instances may arise where explanations for misinformation conflict due to different interpretations brought by these factors. Understanding the dynamics between different explanations is crucial, as it can guide further research on effectively presenting explanations to enhance user comprehension and decision-making. For instance, if it is found that conflicting explanations lead to reduced detection accuracy due to user confusion, the research could then focus on finding an effective way of presenting these explanations (e.g., implementing a sequential presentation strategy where the primary explanation is provided first, followed by the other explanation design upon user request for further detail). On the other hand, if differences between content and social explanations are found to enhance detection ability by fostering critical thinking \cite{lodge2018understanding, schwind2012reducing}, future efforts could be directed toward designing methods for seamlessly presenting these conflicting explanations.

Motivated by the above research gaps, we conducted two online crowdsourcing-based studies to assess how different types of explanations influence users' abilities to detect misinformation and enhance their overall experience. We evaluated four explanation types in \textbf{Study 1}, using Prolific: content, social, aligned (where content and social explanations agree), and misaligned (where they conflict), within the context of a COVID-19 misinformation detection task. Our findings indicated that aligned explanations significantly improved both detection accuracy and user experience compared to content and social explanations. In contrast, misaligned explanations proved no more effective than providing no explanation at all in enhancing detection accuracy. \textbf{Study 2} aimed to confirm and extend these findings to both COVID-19 and political misinformation, using a different participant pool from Amazon Mechanical Turk (MTurk). We observed that aligned explanations consistently enhanced misinformation detection accuracy across these domains, reaffirming the results from Study 1. Additionally, the effectiveness of explanations varied depending on the topic domain and presentation order.  In both studies, we explored the nuanced ways in which participants make decisions, particularly when presented with misaligned explanations. Specifically, our qualitative analysis showed that misaligned explanations can foster participants to either reinforce their existing beliefs or engage in critical thinking by scrutinizing the explanations and seeking for more evidence. 

\rblue{Examining the effects of content-based and social explanations on users' ability to detect misinformation is crucial, as these cues are not always available or aligned in real-world settings \cite{jeronimo2019fake, zhang2018fauxbuster}. For example, socio-contextual cues like engagement metrics or credibility indicators may be sparse or entirely absent in emerging misinformation scenarios—such as those involving newly created accounts or unverified sources—or unreliable when socio-contextual information is intentionally distorted to mislead the audience (e.g., \cite{jeronimo2019fake, shao2018spread}). In all such cases, detection models should primarily rely on content features. Conversely, when a text is extremely short, missing altogether (as in images or memes), or lacks meaningful linguistic patterns, content analysis alone may be insufficient, making socio-contextual cues more critical for credibility assessment (e.g., \cite{zhang2018fauxbuster, li2023assessing}). Thus, when detection systems can only leverage one type of cue, it is essential to assess how effectively content-based and social explanations support users’ ability to detect misinformation when used in isolation. Furthermore, even when both types of cues are available, they may not always align, as content-based and social explanations highlight different aspects of misinformation (see Figure 2 for an example). This potential misalignment raises important questions about their individual and combined influence: Should users be exposed to both explanations to reveal contradictions, or is it more effective to present only the explanation that aligns with the model’s final verdict? On the one hand, misaligned explanation may prompt users to critically evaluate each of the explanation, which can be especially beneficial when the model's prediction is incorrect. On the other hand, misaligned explanation can also create confusion or cognitive overload due to the conflicting information being presented, making it unclear whether users benefit from seeing both explanations together. If the latter, would it be more beneficial to rely on on just one type of explanation (e.g., content explanation) to reduce cognitive strain, or would doing so risk overlooking critical information or unique effect that the other explanation (e.g., social explanation) might offer? To address these questions, our study systematically investigate how content and social explanations—both independently and in combination—affect users’ misinformation detection accuracy and overall experience.} We summarize our work's contributions as follows:

\begin{itemize}
    \item We conducted the \textit{first} empirical study that compared the effectiveness of different explanation types—content, social, aligned, and misaligned—on users' misinformation detection ability and their user experience. Aligned explanations were found to significantly improve detection accuracy and user experience, while misaligned explanations showed no notable advantage over having no explanation 
    in terms of detection accuracy.   
    \item We identified distinct differences in how the order of explanation types influences misinformation detection accuracy across domains. Specifically, in the COVID-19 domain, the effectiveness of aligned and misaligned explanations significantly depended on which type of explanation was presented first. \rzre{However, this interaction effect was not found in the political domain.}
    \item Based on our findings, we offer design recommendations for developing more effective explanations of misinformation. Specifically, we suggest to: 1) prioritize alignment between content and social explanations to improve both accuracy and user experience, 2) explore ways to mitigate the potential risks of misaligned explanations in reinforcing confirmation bias, and 3) tailor explanations to specific domains, as their effectiveness can vary depending on the context. 
\end{itemize}



%

%% file: 2related.tex
\section{Related Work}
\label{sec:related}

\subsection{Two Types of Misinformation Explanations: Content Explanation vs. Social Explanation} 
Our work is directly motivated by the existing gap between research in misinformation detection and XAI. Misinformation detection models have achieved high accuracy by leveraging both the content cues (e.g., syntax and semantic cues) and social-contextual cues (e.g., user attributes and information dissemination context) \cite{shu2019beyond, shu2017fake} (see Table \ref{tab:misinfolit} for examples). Each type of cues is known to provide a unique perspective in identifying misinformation. For example, content cues offer insights into the textual structure and logical coherence of information, aiding in identifying patterns of misinformation based on language use \cite{choudhary2021linguistic,carrasco2022fingerprints,jeronimo2019fake,vieira2020analysis}. Socio-contextual cues, on the other hand, consider user attributes \cite{cui2022meta,shu2019role,long2017fake} and context of information dissemination \cite{vosoughi2018spread,zhang2018fauxbuster}, which are crucial in understanding the intent and reliability behind the information. Leveraging these cues, misinformation detection models have shown high detection accuracy, with some studies reporting over 90\% accuracy even within five minutes of news dissemination \cite{ghosh2023catching, liu2018early}. 

\begin{table*}[htb!]
  \footnotesize
  \centering
  \caption{Examples of Content and Socio-Contextual Cues Identified in Misinformation Detection Literature}

  \resizebox{\textwidth}{!}{
    \begin{tabular}{l | l | l}
    \toprule[1.5pt]
    \textbf{Content cues} & \textbf{Indicators of Truth} & \textbf{Indicators of Misinformation}\\
    \midrule[1pt]
    Syntactic Cues & Active voice, specific pronouns. & Passive voice, vague or excessive personal pronouns. \\
    Semantic Cues & Neutral tone, factual or objective language. & Emotional tone, subjective or opinionated language.\\
    \midrule[1.5pt]
    \textbf{Socio-contextual cues} & \textbf{Indicators of Truth} & \textbf{Indicators of Misinformation}\\
    \midrule[1pt]
    User Attributes & High domain expertise, track record of reliability. & Low domain expertise, track record of misinformation. \\
    Information Dissemination Context  
    & High-credibility sources with high fact-checking. & Low-credibility sources with little to no fact-checking. \\
    \bottomrule[1.5pt]
    \end{tabular}
  }
\label{tab:misinfolit}
\end{table*} 


Despite the significant roles played by both types of cues, current XAI methods predominantly focus on content cues, often neglecting socio-contextual cues. This research gap is well-documented by \citet{gong2024integrating}, where they proposed a vision for incorporating social-contextual cues into an explanatory approach, which they termed ``social explanation.'' While their work highlights the potential benefits of social explanations in enhancing user understanding and trust in misinformation detection systems, they have primarily provided a conceptual framework rather than concrete developments. Our work significantly extends their work in two ways. First, we develop social explanations and conduct an empirical evaluation of their effectiveness. Second, we explore the intricate interplay between content and social explanations by examining scenarios in which both types of explanations are presented, either in alignment or misalignment with one another, as well as the impact of their presentation order. In addition, a few studies have presented both content and social explanations to users to assess their impact on misinformation detection \cite{seo2024reliability, yang2019xfake}. For example, \citet{yang2019xfake} introduced the XFake system, an explainable fake news detector designed to aid users in assessing news credibility by analyzing linguistic features (content cues) and surrounding social context and speaker information (socio-contextual cues) of the claim. However, while their system demonstrates how explaining both types of cues aids in understanding and predicting the model's decision-making process,
it primarily assesses the \textit{collective} impact rather than investigating potential interaction effects between these distinct types of cues. 
Similarly, \citet{seo2024reliability} explored AI explanations that included content, source, and social credibility, but focused on how the framing of the explanations (i.e., positive or negative) affects users' detection of misinformation, without investigating the potential interaction effects between these different types of explanations.

\textbf{This work.} Building on gaps identified, we provide a detailed analysis of how content and social explanations, both individually and in combination, influence misinformation detection and user experience. Our study demonstrates that the alignment of these explanations significantly impact misinformation detection, with effectiveness varying across domains and presentation order—a novel finding that has yet been explored within XAI research on misinformation.



\subsection{AI Explanations in Decision Making and Misinformation Detection}
Our work is also closely related to research on the effectiveness of AI explanations in decision-making. Various studies within the Human-Computer Interaction (HCI) and Computer-Supported Cooperative Work (CSCW) communities have explored how AI explanations affect human decision-making, but their findings are inconsistent. Prior work indicates that explanations improve understanding of AI models, which assist users in making informed decisions \cite{cheng2019explaining}. However, other studies found that explanations might lead to either over- or under-reliance in AI systems, negatively impacting task performance and user satisfaction \cite{jacobs2021designing,springer2019progressive}. A recent study by \citet{vasconcelos2023explanations} suggests the mixed results could be due to individuals weighing the cognitive effort required to understand and process an explanation against the expected benefits. In simpler terms, if an explanation is complex, users might perceive the cognitive effort as surpassing the benefits of comprehending the explanation, resulting in not fully engaging with AI decisions. Consequently, the XAI community is moving towards more user-friendly explanation methods \cite{liao2021human}. A significant development in this direction is the use of natural language explanations, which have been shown to improve user comprehension by clearly articulating the AI system's rationale \cite{kou2022crowd} as well as resonating with how humans understand and generate explanations \cite{ehsan2024xai}. As a result, studies have increasingly explored the use of natural language explanations to assist users in various AI-assisted decision-making contexts, such as image classification \cite{morrison2024impact}, question answering systems \cite{pafla2024unraveling}, recommendation systems \cite{balog2023measuring}, clinical decision support \cite{panigutti2022understanding}, and data annotation tasks \cite{wang2024human}.  


Despite the broader trend towards using natural language explanations for AI systems, their application in misinformation detection remains limited. Some studies have focused on adding explanations to warning labels, such as 
by adding details about the source of the flag (e.g., ``This claim has been flagged by independent fact-checkers.'') \cite{horne2024does}, explaining the generation process (e.g., ``The detection system relies on trained users who report on the veracity of news posts.'') \cite{epstein2022explanations}, or providing the fact-based explanation (e.g., ``There is no evidence to support this claim.'') \cite{barman2023does}. While these explanations were shown to enhance user understanding and detection accuracy more effectively than warnings alone \cite{epstein2022explanations}, they provide minimal insight into \textit{why} the claim was flagged as misinformation, such as the specific factors or reasoning behind the decision. 
Other work, similar to our approach, has aimed to provide explanations using content or socio-contextual cues \cite{yang2019xfake, seo2024reliability, mohseni2021machine, purificato2022tell}. However, to our knowledge, none of these studies have explored natural language explanations; instead, they have primarily focused on visual formats such as bar charts \cite{seo2024reliability, mohseni2021machine}, pie charts \cite{purificato2022tell}, graph triples \cite{kou2022hc, shang2022duo}, and customized feature impact visualizations \cite{yang2019xfake} to present the key features that contribute to the model's decision-making process. While visualizations have advantages in that they are intuitive, they often oversimplify and abstract complex data by condensing detailed information into simplified graphics or charts \cite{jambor2024ten}. This oversimplification can obscure important details and contexts, making it harder for users to fully understand how the data supports the model’s decisions, particularly for those unfamiliar with AI \cite{ehsan2019automated, ehsan2021explainable}.  



\textbf{This work.} We build upon and extend recent trends in XAI research by designing content and social explanations of misinformation using natural language. Our work is among the first to explore how content and socio-contextual cues can be integrated through natural language explanations within the field of XAI for misinformation detection. To this end, we developed a method using a large language model to generate content and social explanations from relevant cues, where the resulting explanations have been shown to enhance users' ability to detect misinformation.

%% file: 3gpt.tex
\section{Dataset and Explanation Generation Process}
\label{sec:gpt}
This section details our methodology for generating AI explanations of misinformation, including our dataset selection and validation processes, followed by the explanation generation process. 

\subsection{Dataset and Ground Truth Label}
The claims analyzed in this study were sourced from PolitiFact, a reputable third-party fact-checking organization.\footnote{\url{https://www.politifact.com}} While there are various fact-checking services and datasets available, different services provide distinct socio-contextual cues, which can complicate the generation of coherent social explanations. Therefore, we exclusively chose to collect data from PolitiFact among others for several reasons: First, it is one of the most widely cited sources in misinformation detection literature, with some papers exclusively relying on using data from Politifact \cite{mosleh2022measuring, linder2021level, yang2019xfake}. Second, PolitiFact is confirmed by the International Fact-Checking Network (IFCN)\footnote{\url{https://www.poynter.org/ifcn/about-ifcn/}} to be non-partisan and fair fact-checking organization. Third, PolitiFact provides various socio-contextual cues, such as speaker information (e.g., speaker name, political affiliation, credibility history, job title) and the context of the claim. Finally, PolitiFact provides professional justifications and includes URL links to sources, which facilitates the verification of claims. Specifically, to mitigate potential inaccuracies from professional fact-checkers, we meticulously reviewed and cross-validated the accuracy of each claim's label. This process involved corroborations claims from other reputable sources such as Snopes\footnote{\url{https://www.snopes.com}}, Factcheck.org\footnote{\url{https://www.factcheck.org}}, and additional reliable online resources. For those that were not verified by other sources, we carefully read the justifications from PolitiFact and checked whether the explanations were backed up by concrete evidences (e.g., statistics, expert consultations). Only those claims that were supported by at least one other credible source or were justifiable were included in our dataset. 

Among the various domains provided by PolitiFact, we focused on COVID-19 (Study 1 and 2) and Politcs (Study 2) for their societal impact and unique misinformation characteristics \cite{muhammed2022disaster}. COVID-19 misinformation often incites panic, such as baseless reports of vaccine-related child deaths, or spreads false hope, like rumors of free medicine distribution, whereas political misinformation typically aims to influence public opinion or affect policy decisions \cite{muhammed2022disaster}. To simplify the analysis, we only collected claims labeled as ``True'' and ``False,'' from the six labels categorized by PolitiFact\footnote{We excluded categories representing partial truths (``Mostly True,'' ``Half True,'' ``Mostly False'') to focus on clear dichotomies, thereby reducing ambiguity in our analysis. We also omitted ``Pants on Fire'' claims to avoid skewing our evaluation, as their extreme falsehood could make them easier to detect.}: True, Mostly True, Half True, Mostly False, False, and Pants on Fire. For COVID-19, we manually collected the claims that contained socio-contextual cues from the ``Coronavirus'' category under the ``Issues'' menu on the PolitiFact website, resulting in 43 true claims and the 72 most recent false COVID-19 claims. For the political claims, we randomly selected 61 true and 61 false claims from the training data in the LIAR dataset \cite{wang2017liar}, which is sourced from PolitiFact. 

\subsection{Explanation Generation Process}
\label{sec:generation}

\begin{figure*}[htb!] 
    \centering
    \includegraphics[height=6cm, width=0.95\textwidth]{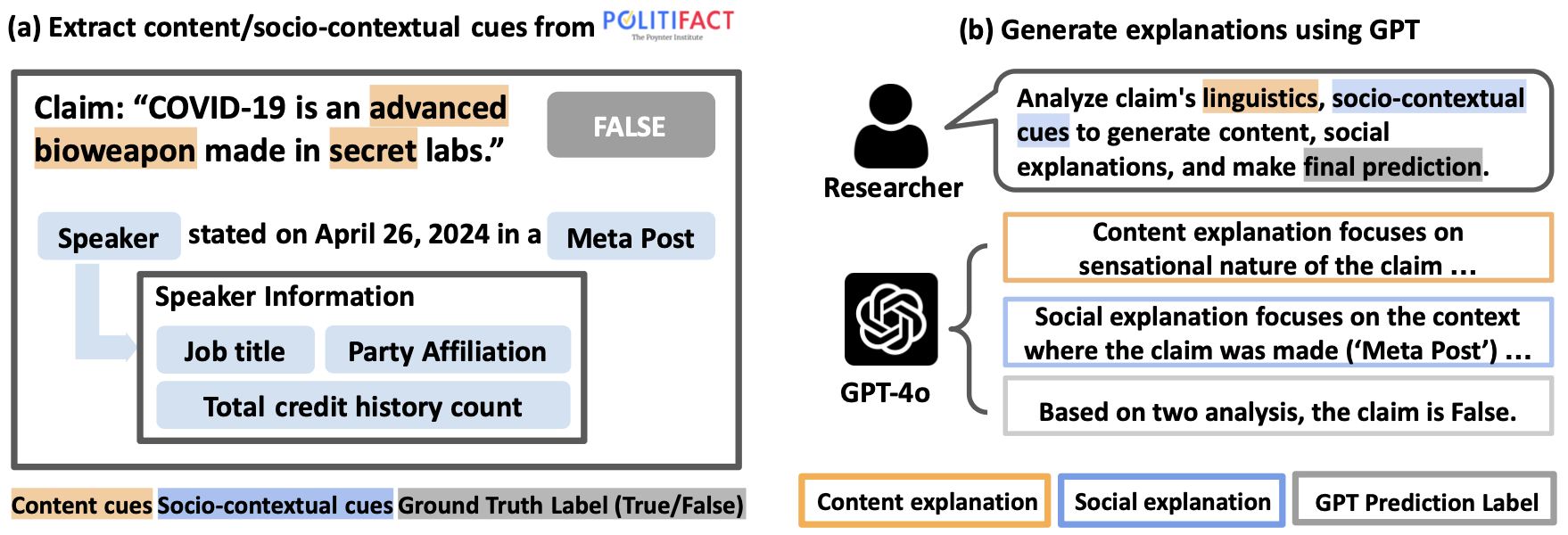}
    \caption{Overview of the Explanation Generation Process Using GPT.}   
    \label{fig:exp}
\end{figure*}

To generate content and social explanations, we extracted content and socio-contextual cues from each claim, as well as its corresponding ground truth label from our dataset (Figure \ref{fig:exp}(a)). Using the extracted content and socio-contextual cues, we generated corresponding content and social explanations using GPT-4o-2024-08-06, the latest GPT model available at the time of the experiment. \rblue{The selection of content (linguistic) and socio-contextual cues was guided by their importance in misinformation research \cite{shu2017fake} and their availability in our dataset (particularly for social explanations). Specifically, for content explanations, we instructed GPT-4o to analyze \textit{syntactic, semantic, and structural} cues (Figure \ref{fig:exp}(a), orange section), which are well-documented indicators of deceptive and misleading language in misinformation research \cite{choudhary2021linguistic, shu2017fake}. Based on this analysis, GPT-4o then generated a content explanation that highlights key linguistic patterns contributing to the claim's credibility (Figure \ref{fig:exp}(b), orange box). For social explanations, we incorporated socio-contextual cues such as \textit{speaker attributes (i.e., speaker name, job title, party affiliation, and credibility history)} and \textit{the context of the claim dissemination} (Figure \ref{fig:exp}(b), blue section), both of which are commonly discussed in misinformation studies \cite{long2017fake, shu2019role}. Although engagement-driven cues (e.g., likes, shares, retweets) and network propagation information \cite{bo2023will, guo2018rumor, vosoughi2018spread} have been explored in misinformation research, our dataset lacked such social network metadata, preventing their inclusion in our social explanations. Similar to content explanations, we tasked GPT-4o with analyzing the available socio-contextual cues and generating an explanation that explicitly links these cues to their role in shaping the credibility of the claim (Figure \ref{fig:exp}(b), blue box). For both content and social explanations, we tasked GPT-4o with generating two to three detailed sentences based solely on the provided cues for conciseness. In addition, rather than explicitly specifying how each cue should indicate misinformation, we leveraged GPT-4o’s pre-trained knowledge and reasoning abilities \cite{petroni2019language, radford2019language}, allowing it to interpret and generate explanations based on the provided cues. This approach provided greater flexibility in identifying relevant linguistic cues compared to prior work, particularly for content explanations \cite{gravanis2019behind, kumar2024feature}, as it allowed us to flexibly derive linguistic patterns indicative of misinformation within syntactic, semantic, and structural dimensions, rather than being restricted to a fixed linguistic cue list.} 

\rblue{Furthermore, since our study aimed to capture instances of misalignment between explanations, we prompted GPT-4o to analyze and generate content-based and social explanations \textit{separately}. This decision was based on our observation during our experiment with various prompting methods for explanation generation that when the model generates both explanations together, it tends to align its reasoning across explanation types, resulting in an over-distribution of aligned explanations.}
Finally, we asked GPT-4o to provide the final verdict by considering both the content and social analyses previously conducted \rblue{individually} (Figure \ref{fig:exp}(b, grey part)). This entire process was applied to all claims in our dataset. (\mt{See Appendix A for the design rationale} and Appendix \mt{B} for the prompting methods used in explanation generation.)

After generating explanations for each claim, we conducted a thorough verification to ensure the explanation quality. Recognizing the inherent variability of GPT \cite{liao2023ai}, we repeated the misinformation detection process outlined in the previous paragraph three times for each claim.\footnote{The reliability across trials was 98.25\%, showing the high consistency of final predictions across three trials. The overall detection accuracy averaged 67.45\%, with aligned claims at 43.67\% and misaligned claims at 87.22\%.} Only those claims where GPT's final prediction (Figure \ref{fig:exp}(b) grey part) was consistent across all three trials were selected. \rblue{After selecting claims based on prediction consistency, we examined the relationship between the generated content-based and social explanations to determine whether they were aligned or misaligned. An explanation was classified as aligned if both explanations supported the same final prediction—either both indicating the claim was true or both indicating it was false. Conversely, an explanation was classified as misaligned if one suggested the claim was true while the other suggested it was false. Based on this classification, we ensured the final dataset contained an equal number of aligned and misaligned explanations. Furthermore, for misaligned explanations, where either the content-based or social-based explanation aligned with the model’s prediction while the other contradicted it, we maintained a balanced distribution by alternating which explanation type aligned with the model’s final prediction. This approach systematically introduced misalignment in a controlled manner, preventing bias toward any particular explanation type in relation to the model’s decision.} To construct the final dataset for the experiment, we randomly selected 24 claims from the balanced dataset, ensuring an equal distribution of ``true'' and ``false'' verdicts as well as equal representation across the COVID-19 and political domains. For each of these 24 claims, we selected the first content-based and social explanations generated from the three trials to present to participants. Finally, to maintain high explanation accuracy, our team—comprising an expert in misinformation detection with extensive peer-reviewed publications—conducted a thorough review of the explanations for the selected claims. This review, based on their expertise and guided by misinformation detection literature (Table \ref{tab:misinfolit}), ensured the explanations accurately reflected the decision-making process in misinformation detection models. (See Appendix \mt{C} for examples of claims and explanations used in the experiment.) 

\rblue{Importantly, by leveraging GPT-4o to generate explicit, natural language explanations, our approach distinguishes itself from prior methods that primarily rely on numerical and visualization-based explanations—such as feature distributions and hierarchical dependency graphs \cite{mohseni2021machine, yang2019xfake}—to convey linguistic and socio-contextual cues that indicate misinformation. In other words, by replacing abstract numerical values and complex visual structures with explanations that articulate how linguistic and socio-contextual cues lead to a claim’s veracity, our approach reduces the cognitive effort required for users to interpret complex representations \cite{ehsan2019automated, ehsan2021explainable}.}



%% file: 3study1.tex
\section{Study 1: Effect of Explanations on Detection Accuracy and User Experience}
\label{sec:study1}
The goal of Study 1 was to assess the effectiveness of different misinformation explanation \rzre{type}s on detection accuracy and user experience. To this end, we conducted a \textit{between-subjects study} where participants were exposed to one of four explanations—content, social, aligned, or misaligned—while engaging in a \textit{COVID-19 } misinformation detection task. 

\subsection{Study Conditions}
\begin{figure}[htb!]
    \includegraphics[width=0.6\columnwidth]{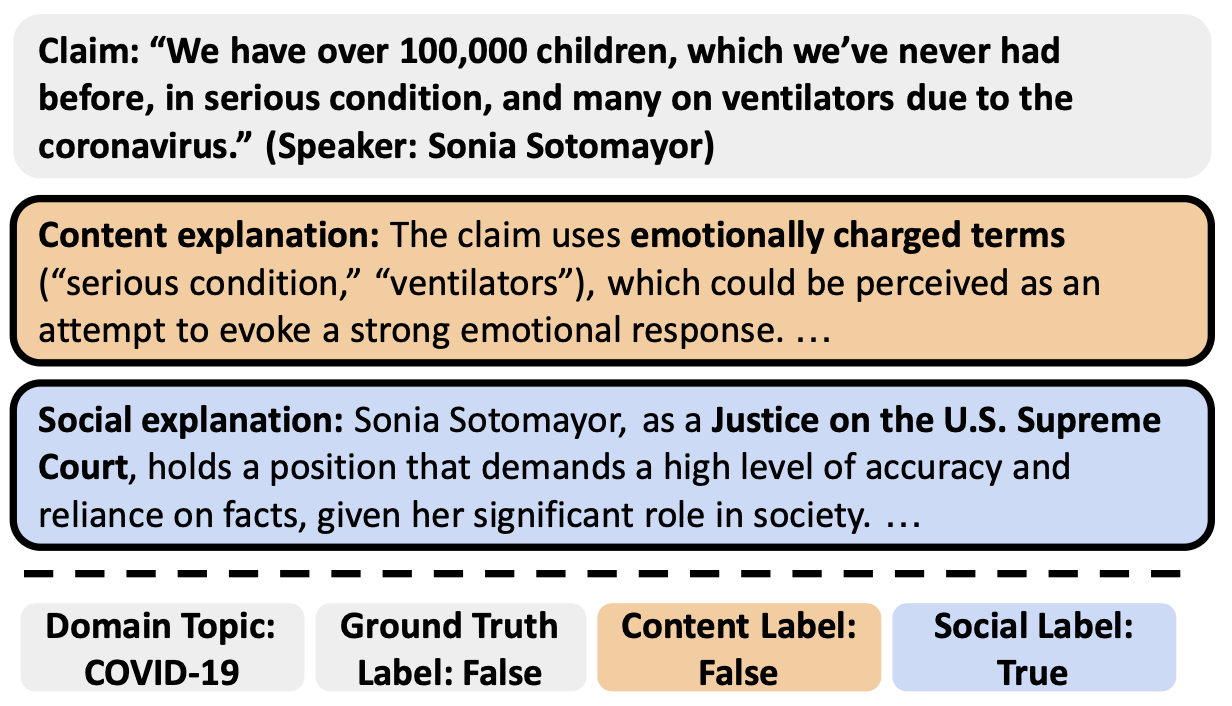}
    \caption{Examples of AI Explanations Shown to Participants. Participants in the content explanation condition were shown explanations in an orange box, while participants in the social explanation were shown explanations in the blue box. The combined explanation conditions (i.e., aligned and misaligned explanation conditions) were shown both sets of explanations in orange and blue boxes, respectively. \rzre{This example represents an ``misaligned'' explanation condition, where the content explanation labels the claim as false, while the social explanation labels it as true. Note: The boxes below the dashed line (Domain Topic, Ground Truth Label, Content Label, Social Label) were not shown to participants during the experiment.}}
    

    \label{fig:studydesign}
\end{figure}

This study explores five conditions with different misinformation explanations: no-explanation (control condition), content, social, and combined explanations. The combined explanation is further divided into two scenarios: one where the content and social explanations are in agreement (aligned) and another where these explanations contradict each other (misaligned). To minimize any ordering effects in the combined explanations, the sequence in which content and social explanations were presented to participants was randomized. Examples of how explanations are presented can be seen in Figure \ref{fig:studydesign}.

\vspace{0.15cm}

    \textbf{(1) No explanation (Control)}: The AI system does not provide any prediction or explanation of any forms at all. 

    \textbf{(2) Content Explanation} (Figure \ref{fig:studydesign} (Orange box)): The AI system provides an explanation for why the statement could be correct or misleading, detailing content cues,  such as 
    syntax and semantics. 

   \textbf{(3) Social Explanation} (Figure \ref{fig:studydesign} (Blue box)): The AI system provides an explanation for why the statement could be correct or misleading, detailing socio-contextual cues, such as user attributes and the context in which the statement was made or disseminated. 

    \textbf{(4) Aligned Explanation}: The AI system provides an explanation for why the statement could be correct or misleading by 
    providing both content and social explanations 
    that are in agreement.
    
    \textbf{(5) Misaligned Explanation} (Figure \ref{fig:studydesign} (Orange and Blue box)): The AI system provides an explanation for why the statement could be correct or misleading by 
    providing both content and social explanations 
    that are in disagreement.

\vspace{0.15cm}

\rblue{While misinformation detection systems typically provide a final verdict along with an explanation, we intentionally excluded the verdict in our study to isolate the impact of explanations themselves. Including a verdict could influence participants' reasoning, making them more likely to accept the system’s decision rather than critically evaluating the explanations on their own \cite{gajos2022people}, especially when the task is cognitively demanding  \cite{vasconcelos2023explanations}. By omitting the verdict, we ensure that participants independently assess the explanations, allowing us to better understand how different types of explanations affect user judgment and overall experience--a key focus of this study.}


\subsection{Participants}
A total of 213 participants from Prolific took part in the experiment and passed the screening test\footnote{To ensure high data quality, we included a screening question to confirm that participants had read the explanations. Those in the content or social explanation conditions indicated whether the explanation classified the claim as true or false, while participants in the combined explanation conditions (i.e., aligned and misaligned explanation conditions) assessed whether the content and social explanations were consistent or inconsistent.}. All participants were U.S. residents, native English speakers, and met the eligibility criteria we set to ensure data quality: a minimum of 1,000 previous submissions and an approval rate of at least 95\%. The participant's age ranged from 22 to 85 years (mean = 46.02, median = 44.5), with a gender distribution of 56.02\% male, 43.52\% female, and 0.46\% preferring not to disclose. \rzre{Based on our power analysis for identifying small effects (power level = 0.8, alpha level = 0.05), at least 40 participants were needed per each condition. To account for potential dropouts, we recruited 50 participants per condition. The final participant numbers after screening test per condition were as follows:} Control (N=44), Content (N=40), Social (N=43), Aligned (N=41), and Misaligned (N=45).

\subsection{Procedure}
Participants in our study went through four phases: 1) a pre-survey; 2) the main evaluation task; 3) a post-survey; and 4) a debriefing session. Participants first provided their consent and completed a pre-survey that included demographic questions, such as their gender and age. In the main \rzre{evaluation} task, participants were shown a claim and asked to assess its veracity. Then, depending on their assigned condition, participants in experimental conditions received different types of explanations (i.e., content, social, aligned or misaligned explanations) and were asked to re-assess its veracity. This process was repeated for a total of eight claims. In the control condition, participants did not receive any explanation and were only asked to evaluate the claim's veracity. Following the main task, a post-survey was conducted to gauge user experiences with the AI explanations, focusing on perceived usefulness and participants' understanding of AI. The study concluded with a comprehensive debriefing session where the ground truth labels of the evaluated claims were transparently revealed to ensure participants do not leave with incorrect information about the claims. We also discussed the potential limitations and imperfections of the AI-generated explanations provided during the study. All study procedures received approval from the Institutional Review Boards of the authors' institutions.

\subsection{Measurements and Analysis Methods}
\subsubsection{Measurements}
To evaluate the effectiveness of AI explanations, we measured four metrics, which are grouped into two categories: Detection Accuracy (Standard Misinformation Detection Accuracy and Weighted Misinformation Detection Accuracy) and User Experience (Perceived Usefulness of Explanation, Perceived Understanding of AI). The measurements from our study are summarized in Table \ref{tab:my_label}.

\begin{table*}[htb!]
    \caption{Summary of Measurements}
    \centering
    \footnotesize
    \begin{tabular}{c l l}
    \toprule
    \textbf{Measurement}& \textbf{Metric} & \textbf{Scale}\\
    \toprule
    I &Standard Misinformation Detection Accuracy (SMDA) & 1-100\\
    \midrule
    II &Weighted Misinformation Detection Accuracy (WMDA) & 1-100\\
    \midrule
    \multirow{4}{*}{III} & - Helpfulness: ``I find the [content/social] explanation helpful for making decisions.''  & \\
    & - Ease of Task: ``Overall, the [content/social] explanation made the task easier.'' & 7-point Scale:\\
    & - Confidence: ``If I want to assess the veracity of a claim, I would feel comfortable using & 1 = ``Not at all''\\
    & the [content/social] explanation to help me find evidence supporting or refuting it.''&  to\\
    \cmidrule{1-2}
    \multirow{2}{*}{IV} & ``After reading the [content/social] explanation, I have gained a clear understanding of& 7 = ``Extremely''\\
    & how AI determines the claim veracity.''&\\
    \toprule
    \multicolumn{3}{l}{\footnotesize 
    I, II: Misinformation Detection Accuracy. 
    III: Perceived Usefulness of Explanation.
    IV: Perceived Understanding of AI.}
    \end{tabular}
    \label{tab:my_label}
\end{table*}


\subsubsection{Misinformation Detection Accuracy} Misinformation detection accuracy was calculated using two metrics: Standard misinformation detection accuracy (hereafter, ``SMDA'') and Weighted misinformation detection accuracy (hereafter, ``WMDA''). SMDA was calculated based on the binary accuracy rating (``Correct''/``Incorrect''), which simply measures whether the participant's response was accurate. The WMDA provides a more nuanced measure by incorporating both the binary accuracy and the participant's confidence in their judgment. Following the method suggested by \citet{danry2023don}, we calculated the WMDA score by considering for both the binary accuracy rating and a confidence level on a scale of 1-7 (1 = ``Not confident at all'' and 7 = ``Very confident''). The weighted score is then scaled to a range of 0 to 100, with higher WMDA scores indicating better misinformation detection capability. 

\subsubsection{User Experience of Explanation} To evaluate the user experience of explanations, we evaluated the following metrics: the perceived usefulness of the explanations and perceived understanding of AI. The perceived usefulness was measured by calculating the average of the individual 1-7 scale scores for helpfulness, ease of task, and confidence. The questions related to these measurements were brought from \cite{lai2023selective}, and they were adjusted to better align with the context of our study on misinformation explanation.

\subsubsection{Analysis} We analyzed the data using either an ANOVA or a Kruskal-Wallis test across the five conditions for each of the four metrics, depending on whether the data met the assumptions of normality and homogeneity of variances. If these assumptions were met, we used a one-way ANOVA, followed by Tukey’s HSD for post-hoc analysis when significance difference was shown. If not, we used a Kruskal-Wallis test with a Dunn’s test for post-hoc comparisons.

\subsection{Results}
\label{sec:finding}
Below, we report our findings on the effectiveness of misinformation explanations in enhancing users' ability to detect misinformation and improving their overall user experience. Since none of the datasets for the four metrics met parametric assumptions, we applied the Kruskal-Wallis test across all conditions, followed by Dunn’s post-hoc test with Bonferroni correction for significant statistical differences. The descriptive statistics for misinformation detection accuracy and user experience is presented in Table \ref{tab:misresult}.

\begin{table*}[htb!]
  \centering
  \caption{Descriptive Statistics for Misinformation Detection Accuracy and User Experience Across Study 1 Conditions.}
  \begin{tabular}{rcccc}
    \toprule
    \multirow{2}{*}{Conditions} & \multicolumn{2}{c}{\textbf{Detection Accuracy}} & \multicolumn{2}{c}{\textbf{User Experience}} \\
    \cmidrule(r){2-3} \cmidrule(l){4-5}
    & SMDA (\%) & WMDA (\%) & Perceived Usefulness (1-7) & Understanding of AI (1-7) \\
    \midrule
    Control                & 57.67 (18.44) & 40.69 (18.03) & - & -  \\
    Content                & 73.06 (13.03) & 60.14 (15.80) & 4.39 (1.23) & 4.00 (1.14) \\
    Social                & 70.64 (19.01) & 66.95 (20.77)  & 4.55 (1.21) & 3.70 (0.78)  \\
    \textbf{Aligned}    & \textbf{85.06 (5.02)} & \textbf{82.84 (5.49)} & \textbf{5.74 (0.84)}& \textbf{4.66 (1.14)} \\
    Misaligned & 60.56 (9.38) & 44.45 (11.39) & 4.84 (1.27) & 3.67 (1.27) \\
    \bottomrule
  \end{tabular}
  \label{tab:misresult}
  \smallskip
  \\
    \begin{flushleft}
    \tt\small{Note: Values are presented as means, with standard deviations in parentheses. SMDA stands for Standard Misinformation Detection Accuracy, while WDMA stands for Weighted Misinformation Detection Accuracy.}
    \end{flushleft}
\end{table*}


\subsubsection{Effects on Misinformation Detection Accuracy} 
In our study, we evaluated the effectiveness of different explanation types on misinformation detection, using both SMDA and WMDA as measures. Kruskal-Wallis tests conducted for both metrics demonstrated significant differences across the five experimental conditions: control, content, social, aligned, and misaligned (SMDA: H(4) = 180.10, \( p < 0.001 \); WMDA: H(4) = 176.83, \( p < 0.001 \)). Subsequent post-hoc Dunn’s tests revealed that, for both metrics, the aligned condition (mean = 85.06) significantly outperformed the content (mean = 73.06), social (mean = 70.64), misaligned (mean = 60.56), and control (mean = 57.67) conditions (\( p < 0.001 \) for all comparisons). Both content and social explanations were also found to significantly enhance misinformation detection ability compared to the misaligned and control conditions (\( p < 0.001 \) for all comparisons). However, no significant differences were found between the content and social explanations (SMDA: \( p = 0.209 \); WMDA: \( p = 0.140 \)), and between the misaligned and control condition (SMDA: \( p = 0.241 \); WMDA: \( p = 0.140 \)). 

\subsubsection{Effects on User Experience}
Next, we examined how different explanation types impacted user experience by measuring perceived usefulness and understanding of AI. For both metrics, a Kruskal-Wallis test revealed significant differences across the four conditions: content, social, aligned, and misaligned (Perceived usefulness: H(3) = 97.63, \( p < 0.001 \); Understanding of AI: H(3) = 81.374, \( p < 0.001 \)). Post-hoc Dunn’s tests showed that aligned condition (Perceived usefulness: mean = 5.74; Understanding of AI: mean = 4.66) was significantly more effective in both perceived usefulness and understanding of AI compared to content (Perceived usefulness: mean = 4.39; Understanding of AI: mean = 4.00), social (Perceived usefulness: mean = 4.55; Understanding of AI: mean = 3.70), and misaligned (Perceived usefulness: mean = 4.84; Understanding of AI: mean = 3.67) (\( p < 0.001 \) for all comparisons). However, there were notable differences between the conditions for the two metrics. For perceived usefulness, 
misaligned explanation condition was shown to significantly enhance the perceived usefulness compared to content (\( p < 0.001 \)) and social explanation (\( p = 0.027)\) conditions. For understanding of AI, content explanation condition was shown to be more effective compared to misaligned (\( p = 0.021 \)) and social explanation (\( p = 0.032)\) conditions. No significant differences were found between the content and social explanations for perceived usefulness (\( p = 0.199)\), nor between the misaligned and social explanation conditions for understanding of AI (\( p = 0.799 \)).



\vspace{0.25cm}
To summarize, our results suggest that \textbf{\textit{alignment}} between content and social explanations is critical to the effectiveness of misinformation explanations, with aligned explanations significantly enhancing both misinformation detection accuracy and user experience. In contrast, misaligned explanations did not significantly improve detection accuracy over the control, content, and social explanation conditions, although participants perceived them as more useful for making decisions than when presented with either content or social explanations. This discrepancy between the perceived usefulness of misaligned explanations and their actual impact on detection accuracy naturally lead to the question: ``Why might participants perceive misaligned explanations as useful despite their ineffectiveness in enhancing detection accuracy?'' 
\rblue{One hypothesis is that misaligned explanations might reinforce users' initial beliefs due to confirmation bias, where individuals tend to favor information that aligns with their pre-existing views \cite{klayman1995varieties}. Unlike content or social explanations, misaligned explanations inherently contain information that \textit{both} align and misalign with a participant's decisions. Studies indicate that once an initial judgment is made, it influences the processing of subsequent information by altering the significance attributed to new data based on its consistency with the initial decision \cite{talluri2018confirmation, hart2009feeling}. Thus, in the misaligned conditions, it is likely that participants focus on the parts of the explanation that support their pre-existing beliefs, thereby reinforcing these beliefs and perceiving the aligned components as particularly useful \cite{windschitl2013so, russo2008goal}. To investigate this hypothesis, we analyzed whether misaligned explanations reinforced confirmation bias by comparing decision revisions and confidence shifts in the misaligned condition with those in the aligned, content, and social explanation conditions. Specifically, if misaligned explanations reinforced confirmation bias, we would expect participants to be less likely to revise their initial decisions when faced with conflicting explanations and more confident in maintaining their original choice. However, we did not find strong evidence that misaligned explanations \textit{uniquely} reinforced confirmation bias beyond the broader trend of participants generally adhering to their initial decisions across all explanation conditions. Additionally, we examined whether the order in which explanations were presented (e.g., content-first vs. social-first, accurate-first vs. inaccurate-first) influenced decision changes under the misaligned condition to assess whether presentation sequence affected how participants engaged with conflicting information. However, we again found no significant impact of explanation order on decision revision in the misaligned explanation condition. (See Appendix \mt{D} for the full results and interpretations of these further analyses.)}

\subsection{Conclusion of Study 1}
Study 1 demonstrates that aligned explanations significantly enhance users' detection accuracy and overall experience, while misaligned explanations provide no notable advantage compared to content or social explanations, or even when no explanation is provided. Both content and social explanations are helpful in improving users' ability to detect misinformation compared to having no explanation, though they do not differ significantly from each other in their impact on detection accuracy. We also find nuanced impacts of different explanation types on user experience. Notably, while participants perceived misaligned explanations as useful, these explanations did not improve detection accuracy or enhance understanding of AI compared to when content or social explanations were provided alone. \rblue{Our further analysis of the misaligned explanation condition suggests that confirmation bias may play a role in this finding, though other factors could also contribute.}
Additionally, we found that content explanations were more effective in helping participants understand how AI makes decisions compared to misaligned or social explanations, suggesting that different types of explanations have distinct impacts on how people perceive and process information. Altogether, this work addresses a critical gap in the existing literature, which has predominantly focused on content explanations, by examining the comparative effectiveness of content and social explanations and exploring their interactions.

However, this study has several limitations. First, while we quantitatively assess detection accuracy and user experience, we lack qualitative insights into users' subjective experiences with the explanations, which could offer a deeper understanding of their decision making processes. Second, the study’s focus on COVID-19 misinformation raises questions about whether these findings can be generalized to other domains. Third, although the order of explanation presentation (e.g., content-first vs. social-first) could impact participants' decision-making, our current experimental setup—where content and social explanations are shown simultaneously—may not make this order explicit enough for participants to effectively anchor to the initial explanation. 
Acknowledging these limitations, Study 2 address these gaps by incorporating qualitative measures, exploring additional misinformation domains (i.e., politics), and refining the explanation presentation sequence \rzre{to more systematically evaluate the \textit{sequential} influence of different explanation types on misinformation detection.}


%% file: 4study2.tex
\section{Study 2: Effects of Explanation Order and Alignment in Detection Accuracy}
\label{sec:study2}

The Study 2 has two main aims. The first one is to test the robustness and generalizability of the findings from Study 1 while addressing its limitations. The Second one is to to examine how the sequence and coherence between explanation types affect user's misinformation detection ability in both \textit{COVID-19} and \textit{Politics} domains. To this end, we conducted a \textit{2x2 mixed content} study; the between-subjects factor was the type of explanation presented first (either content or social explanation) and the within-subjects factor was the alignment of the second explanation — whether it aligned or misaligned with the initial explanation. We used similar experimental materials and setup as in Study 1, with minor changes detailed in the corresponding sections.



\subsection{Study Conditions}
In Study 2, each participant was assigned to one of two groups based on the between-subjects factor (content or social explanation) and experienced both levels of the within-subjects factor (aligned or misaligned explanations). This setup resulted in five study conditions as follows:

    \textbf{(1) No explanation (Control)}: The AI system does not provide any prediction or explanation of any forms at all.
    
    \textbf{(2) Content-Aligned Explanation}: Participants first received content explanations, followed by social explanations that were aligned with the content explanations.     
    
    \textbf{(3) Content-Misaligned Explanation}: Participants first received content explanations, followed by social explanations that were misaligned with the content explanations. 
    
    \textbf{(4) Social-Aligned Explanation}: Participants first received social explanations, followed by content explanations that were aligned with the social explanations.
    
    \textbf{(5) Social-Misaligned Explanation}: Participants first received social explanations, followed by content explanations that were misaligned with the social explanations.

\subsection{Participants, Procedure, Measurements, and Analysis Methods}
\subsubsection{Participants} 
A total of 231 participants from Amazon Mechanical Turk (MTurk) took part in the experiment and passed the screening test.\footnote{Unlike Study 1, where participants were recruited from Prolific, we recruited participants from MTurk in Study 2 to test the robustness of our results across different platforms. The screening test was the same as the combined explanation (i.e., aligned, misaligned explanation) conditions in Study 1.}
All participants were U.S. residents and their ages ranged from 24 to 59 years (mean=35.14, median=35.00). The gender distribution was 38.41\% female and 61.59\% male. Based on our power analysis for identifying small effects (power level=0.8, alpha level=0.05), at least 34 participants were needed for each condition. The final participant numbers across two different topics were as follows: Control (COVID-19: N=39, Politics: N=34), Content (COVID-19: N=38, Politics: N=41), and Social (COVID-19: N=43, Politics: N=36) explanations.\footnote{In the Content and Social groups, each participant experienced both aligned and misaligned scenarios. For example, the 38 participants in the Content group for COVID-19 received both Content-Aligned and Content-Misaligned explanations.}

\subsubsection{Procedure} 
The procedure in Study 2 was similar to that in Study 1, with three key modifications. First, we removed the pre-decision task, meaning participants no longer made an initial judgment before viewing the explanations. Instead, participants determined whether the claim was true or false \textit{after} seeing the first explanation (e.g., content or social explanation) and then re-evaluated their decision after being presented with a second explanation, which either aligned or misaligned with the first. Second, we omitted the user experience questions from the post-survey to focus more directly on misinformation detection accuracy. Third, we added an open-ended question to gain insights into participants' decision-making processes, specifically to understand how they integrated and reconciled information when content and social explanations misaligned. 

\subsubsection{Measurements and Analysis Methods} Study 2 focused on quantitative measures of misinformation detection accuracy, complemented by qualitative insights into participants' decision-making processes, derived from responses to an open-ended question. \textit{Quantitatively}, a 2x2 Mixed-content ANOVA was conducted to examine the effects of the first explanation type (content vs. social) as a between-subjects factor, and the alignment of the second explanation (aligned vs. misaligned) as a within-subjects factor. Significant findings were further analyzed using Tukey’s HSD for post-hoc comparisons. \textit{Qualitatively}, we examined participants' responses to open-ended questions to gain insights into their interactions with the misaligned explanations. We conducted an inductive thematic analysis \cite{braun2012thematic}, where themes emerged directly from the data rather than being guided by pre-existing categories. Initially, we familiarized ourselves with the data, which allowed patterns and key themes to emerge naturally. We then identified and coded specific segments of participants' responses, focusing on how they processed and reconciled the explanations provided.

\subsection{Quantitative Results: Effects on Misinformation Detection Accuracy}
Below, we present a summary of our findings on detection accuracy for the COVID-19 and Politics domains, analyzed using two metrics: SMDA and WMDA. For SMDA, a mixed ANOVA indicated that alignment was the only factor showing significant differences across both domains, with no main effects of the initial explanation type or interaction effects between explanation type and alignment observed. In contrast, there was a notable difference between the COVID-19 and Politics domain in terms of the WMDA: significant interaction effects between the explanation type and alignment were observed in the COVID-19 domain, but not in the Politics domain.\footnote{The differences between the two metrics might arise because SMDA primarily measures the objective correctness of participants' responses, while WMDA incorporates participants' confidence, which is influenced by subjective perceptions.} 
For brevity, we only report the statistical analysis of the findings for WMDA here and include the SMDA results in the Appendix \mt{E}. The descriptive statistics for both SMDA and WMDA can be found in Table \ref{tab:study2res}.\footnote{Please note that data for the control condition was collected to observe baseline detection accuracy without explanations, and it was not included in the statistical analyses.} 

\begin{table*}
  \centering
  \caption{Descriptive Statistics for Misinformation Detection Accuracy Across Study 2 Conditions.}
  \begin{tabular}{rcccc}
    \toprule
    \multirow{2}{*}{Conditions} & \multicolumn{2}{c}{\textbf{COVID-19}} & \multicolumn{2}{c}{\textbf{Politics}} \\
    \cmidrule(r){2-3} \cmidrule(l){4-5}
    & SMDA (\%) & WMDA (\%) & SMDA (\%) & WMDA (\%) \\
    \midrule
    Control               & 47.69 (10.23) & 28.85 (10.45) & 56.47 (5.97) & 48.82 (7.80) \\
    Content-Aligned       & \textbf{73.68 (6.83)} & 36.84 (12.83) & 70.24 (8.28) & 59.19 (8.71)  \\
    Content-Misligned     & 31.58 (12.38)  & 29.21 (12.89) & 29.27 (12.19) & 16.91 (12.27) \\
    Social-Aligned        & 68.84 (8.77)  & \textbf{69.07 (9.01)} & \textbf{77.78 (4.17)} & \textbf{67.78 (5.07)} \\
    Social-Misaligned     & 35.35 (12.56)  & 17.91 (15.06) & 22.22 (11.36) & 19.07 (11.39) \\
    \bottomrule
  \end{tabular}
    \begin{flushleft}
    \tt\small{Note: Values are presented as means, with standard deviations in parentheses. SMDA stands for Standard Misinformation Detection Accuracy, while WDMA stands for Weighted Misinformation Detection Accuracy.}
    \end{flushleft}
  \label{tab:study2res}
  \smallskip
\end{table*}

\subsubsection{COVID-19} 
A mixed ANOVA revealed no significant main effect of explanation type on detection accuracy (F(1,79) = 0.789, \( p = 0.377 \)), while a significant main effect of alignment was found (F(1,79) = 21.883, \( p < 0.001 \), \(\eta_p^2\) = 0.217). The \textit{interaction effect} between explanation type and alignment was significant (F(1,79) = 10.929, \( p = 0.001 \), \(\eta_p^2\) = 0.122). Specifically, as shown in Table \ref{tab:study2res}, when a social explanation is presented first, the difference in WMDA between cases where the second explanation is aligned or misaligned was substantial (Mean difference between `Social-Aligned' and `Social-Misaligned': 51.16\% in WMDA). In contrast, for content-first explanations, the difference was much smaller (Mean difference between `Content-Aligned' and `Content-Misaligned': 7.63\% in WMDA). This suggests that the type of explanation presented first plays a critical role in how alignment influences detection accuracy, with alignment having a much stronger impact when a social explanation is presented first in the COVID-19 domain.
\subsubsection{Politics} 
\rzre{In contrast to the COVID-19 results, a mixed ANOVA conducted in the Politics domain indicated no significant interaction effect between explanation type and alignment (F(1,70) = 2.261, \( p = 0.137\)). Additionally, there was no significant main effect of explanation type (F(1,70) = 0.102, \( p = 0.919\)), but a significant main effect of alignment was observed (F(1,70) = 114.891, \( p < 0.001\), \(\eta_p^2\) = 0.621), mirroring the COVID-19 results.}

\vspace{0.25cm}

Together, these results highlight two main findings. First, \textit{aligned explanations} consistently prove to be effective in both SMDA and WMDA across the COVID-19 and Politics domains, re-confirming their significance in enhancing the misinformation detection accuracy. Second, there are distinct \textit{differences between the domains} in WMDA: In the COVID-19 domain, there is a significant interaction effect between the type of the initial explanation and whether the subsequent explanation is aligned. This effect is particularly pronounced when the initial explanation is a social one as opposed to a content-based one. \rzre{On the other hand, in the Politics domain, the interaction effect was not found.}
We discuss the potential reasons for the differences between domains in Section \ref{sec:discussion}.

\subsection{Qualitative Results: Participants' Engagement with Misaligned Explanations}
Our qualitative analysis reveals varied interaction patterns with misaligned explanations among participants. Some reinforced their pre-existing beliefs, while others engaged in deep critical analysis, leading to a reassessment of their positions. \rblue{Additionally, participants varied in how they navigated and perceived different types of explanations: while many participants appreciated both content and social explanations, some found social explanations more helpful, others prioritized content explanations, and a few struggled to reconcile contradictions between the two.} Below, we detail \rblue{four} key themes that emerged from our analysis: reinforcement of initial beliefs, critical analysis of explanations, assessment of credibility and evidence, \rblue{and individual differences in engaging with content and social explanations}. 

\subsubsection{Reinforcement of Initial Belief}
Participants often used explanations to confirm or reinforce their pre-existing beliefs, indicative of a potential confirmation bias. For example, one participant reflected, ``My opinion did not change significantly as I read through the explanations provided. Instead, my initial inclination towards considering the claim as true \textit{was reinforced} by the clear cause-and-effect relationship presented in the claim.'' This participant's focus on elements that confirmed their initial belief, while potentially overlooking contradictory evidence, implies how misaligned explanations can sometimes solidify preconceived notions by fostering the selective information processing discussed in Study 1. Another participant used personal research to validate their decision, stating, ``I reviewed the two explanations, .. \textit{Then, I am sure my final choice is always right} because I studied many articles about COVID activities.'' This response highlights how prior knowledge can interact with provided explanations to reinforce one’s decision, potentially leading to unintended consequences when the existing knowledge is flawed or biased. 

\subsubsection{Critical Analysis of Explanations}
On the other hand, some participants used misaligned explanations to engage in critical analysis, scrutinizing the explanations and identifying inconsistencies between them. This scrunity sometimes resulted in heightened skepticism or a shift in opinion based on the strength of the arguments and evidence encountered in the opposing explanations. For example, one participant mentioned, ``Initially, I believed the claim was true, due to my own personal knowledge on the topic. \textit{However, I did consider that the AI is correct} in that the languages is emotionally charged and subjective, which is \textit{why I then felt moderately confident} in my claim, rather than having a higher level of confidence.'' This response shows that the participant reflected on their initial belief but also critically evaluated the AI's explanation, leading to a reassessment of their confidence level. Another participant showed cautious decision-making by weighing different aspects of the explanations before forming a judgment: ``... Initially, I leaned towards Explanation A due to its clear structure, but as I read further, \textit{I found some gaps in its reasoning.} Explanation B, despite its initial disorganization, presented some compelling points that \textit{swayed my opinion}.''

\subsubsection{Assessing credibility and evidence}
Participants also evaluated the credibility of the source and the evidence provided in the explanations. This tendency to assess credibility and evidence was particularly prominent among those who were uncertain about their decision. For example, one participant described how reassessing the speaker's authority altered their perception of a claim: ``I reevaluated my initial impression and \textit{recognized the significance of [Speaker]'s role}, which made it less likely for her to make unverified claims.'' Other participants emphasized the importance of concrete evidence, although they acknowledged that the explanations were persuasive. For example, one participant stated, ``... However, despite the contextual information provided, I still found it challenging to fully trust the claim \textit{due to the lack of concrete evidence} presented within the content analysis.'' Similarly, another participant noted, ``While the neutral tone and quantitative comparisons suggest reliability, the potential bias introduces uncertainty. \textit{The claim's accuracy cannot be definitively determined without access to supporting evidence.}'' These responses reflect a evidence-based approach to evaluating conflicting information, where the credibility of the source and the strength of the evidence heavily influence the acceptance of claims in such scenarios.

\subsubsection{Individual Differences in Engaging with Explanations} \rblue{Finally, we note that participants engaged with conflicting content and social explanations in distinct ways. While many participants valued having access to both explanations, recognizing their complementary nature in providing a more holistic understanding of claims, their decision-making strategies varied when determining which explanation to ultimately follow. For example, some participants prioritized social explanations, using speaker credibility and historical reliability as their primary decision-making factor. One participant reflected, ``Together, they (Content and social explanations) contributed to a nuanced understanding of the claim's context and potential limitations. However, a social explanation was more helpful in arriving at a decision. It affected my decision, although the content explanation raised some doubts regarding my final choice. Additionally, \textit{considering the credibility history has helped me make decisions confidently.}'' Another participant explained how the social explanation helped provide context and ease doubts raised by the content explanation, stating, ``... The absence of specific data sources and reliance on emotionally charged language remained significant considerations that affected my judgment. \textit{However, the social explanation proved to be more helpful in providing context and mitigating some doubts about the claim's credibility. It underscored the expectations associated with the speaker’s role}, which influenced my decision-making process.'' On the other hand, other participants expressed greater skepticism toward speaker credibility, placing more weight on linguistic cues when assessing a claim. One participant noted, ``Even though the speaker is a well-respected figure, \textit{the way the claim is structured still raises red flags}. The AI pointed out the potential oversimplification and ambiguity in the claim's language, which made me doubt its validity.'' Similarly, some participants argued that the structural and linguistic characteristics of a claim is more directly linked to the claim's credibility than external socio-contextual factors, making content explanations more influential in their decision-making. This preference for content explanation is reflected in the following participant response: ``Both explanations provided valuable insights. ... Ultimately, the content explanation was slightly more helpful in evaluating the claim's veracity, as it directly addressed the claim's structure and language, \textit{which I believe are key factors in assessing credibility.}''}

\rblue{Unlike majority of participants who leaned toward either content or social explanations when making final decisions, a few experienced uncertainty when faced with conflicting information, making it difficult to reconcile the two perspectives. One participant expressed reduced confidence after reading both explanations, stating, ``After reading both the content and social explanation, it has made my decision less confident. \textit{Somewhat, both explanations have made it difficult for me to make a decision.}'' Another participant described feeling `stuck' between the explanations: ``\textit{I was very back and forth on this one.} At first glance, I don’t really know what they were talking about, but bragging from a political person is usually not true, so I went with false. The social context made me a little more confident in being false. The content explanation made me a little less confident, and \textit{I still don’t really know.}'' While this response reflects hesitation in decision-making and the challenge of reconciling conflicting perspectives from the misaligned explanations, it also indicates that participants actively considered both explanation rather than immediately defaulting to one.}


\subsection{Conclusion of Study 2}
The findings from Study 2 confirm and extend the results of Study 1 in four ways: First, the results re-confirm those of Study 1, showing that aligned explanations are the most effective, while misaligned explanations do not significantly improve detection accuracy. 
Second, we find that there is a significant interaction effect in the COVID-19 domain, with a larger gap in detection accuracy between aligned and misaligned explanations when social explanations are presented first. This finding highlights the importance of explanation presentation order on users' decision makings, which was previously unexplored in Study 1.
Third, we find that domain topics play a crucial role in shaping the effectiveness of the explanations' sequence and alignment. Unlike in the COVID-19 domain where there was a significant interaction effect between the explanation type and alignment, no such effect was found in the Politics domain. 
Fourth, our qualitative analysis showed that participants employed critical thinking when engaging with misaligned explanations by critically analyzing the provided information and assessing the credibility of the sources. This suggests that, although misaligned explanations did not significantly increase detection accuracy, they facilitated deeper cognitive engagement by encouraging participants to more thoroughly evaluate the explanations\rblue{, which helps explain the increased perceived usefulness of misaligned explanations observed in Study 1}.



%% file: 5discussion.tex
\section{Discussions and Future Work}
\label{sec:discussion}

In this section, we first summarize the benefits and limitations of presenting both the content and social explanations, domain-specific effectiveness of these explanations, \rblue{as well as the ethical considerations of deploying AI-generated explanations for misinformation detection}. Next, we discuss the implications for designing effective AI explanations, \rblue{the theoretical contributions of our work to the CSCW and XAI communities}, and directions for future work. 

\subsection{Discussions}
\subsubsection{Benefits and Limitations of Explanation Alignment in Misinformation Detection}
\label{sec:dis1}
The results of our study suggest clear benefits and limitations when presenting both content and social explanations to users in identifying misinformation. Specifically, across two studies, we found that aligned explanations, where content and social cues agree with each other, were most effective in enhancing misinformation detection accuracy and user experience. On the other hand, the impact of misaligned explanations on users was complex and showed a dual nature—both hindering and enhancing cognitive engagement. In Study 1, misaligned explanations did not enhance detection accuracy compared to conditions where no explanation was provided. This result could be attributed to participants' tendency to adhere to their initial beliefs when faced with contradictory cues, thereby often disregarding parts of the explanations that conflicted with their preconceptions. However, misaligned explanations were not entirely detrimental. For example, we found that participants perceived these explanations as more useful compared to scenarios where only content or social explanations were presented. Study 2's qualitative data further supported this finding, showing that participants employed misaligned explanations as a catalyst for critical thinking, actively seeking additional evidence or meticulously analyzing the explanations to refine their decisions. Together, the varied user reactions to misaligned explanations highlights their potential to engage users in deep cognitive processing, albeit with the risk of reinforcing incorrect pre-existing beliefs and without necessarily leading to improved performance. 

Our findings underscore the critical role of explanation alignment in enhancing users' ability to detect misinformation and contribute to ongoing efforts in the XAI community advocating for the integration of social explanations. Specifically, \citet{gong2024integrating} has proposed incorporating social explanations in XAI frameworks for misinformation detection, motivated by the gap in the XAI research on misinformation, where much of the 
{existing work has focused predominantly on content explanations alone. Similarly, theoretical frameworks have been presented on how socio-contextual information can be integrated into the current XAI framework through the introduction of the concept of ``Social Transparency'' \cite{ehsan2021expanding}. While these works highlight the value of social explanations, our study extends these theoretical arguments in two significant ways. First, we empirically test the inclusion of social explanations and examine how their alignment with content explanations (i.e., aligned vs. misaligned explanations) influences user engagement and decision-making. Second, our findings on the different effectiveness between aligned and misaligned explanations indicate that beyond simply providing both types of explanations, the alignment and presentation order of these explanations are equally, if not more, crucial to their effectiveness. Thus, further work is needed to explore the nuanced ways in which different types of explanations interact with each other and their collective impact on users, in order to better support misinformation detection.

\subsubsection{Domain-Specific Effectiveness of Explanations} 
\label{sec:dis2}
In \textit{Study 2}, we observed that the effectiveness of explanations varied depending on the domain. Specifically, in the COVID-19 domain, when a social explanation was presented first, the difference in detection accuracy between subsequent aligned and misaligned explanations was significantly larger than when a content explanation was presented first. One possible explanation for this variance is the different types of cognitive processing that the explanations may trigger \cite{reinhard2010content}. Although designed to encourage critical analysis of the claim, social explanations, which convey information about the speaker's background and the context in which the claim is disseminated, often function as heuristic cues. For example, these explanations may activate the authority heuristic, where the perceived credibility or expertise of the speaker prompts users to accept the claim without engaging in thorough analysis \cite{mondak1993public, chaiken1994heuristic}. On the other hand, content explanations detail the linguistic features and logical structure of the claim, which can promote analytical processing as they involves an evaluation of the evidence and logic presented \cite{reinhard2010content}. Psychological studies suggest that while individuals engaged in analytical thinking are less influenced by additional information due to their deeper initial processing, those engaged in heuristic thinking are more susceptible to influence \cite{chaiken1994heuristic}, particularly when confronted with information that is misaligned with their preliminary assessments. In light of these studies, we interpret that when individuals are first presented with social explanations in the COVID-19 misinformation detection task, they may engage in heuristic processing, increasing susceptibility to subsequent misaligned explanations. Conversely, when a content explanation is presented first, the explanation can promote analytical processing, making judgments less vulnerable to be influenced from subsequent explanations.

\rzre{In the political domain, we found no significant interaction effect between the initial explanation type (content or social) and the subsequent explanation alignment (aligned or misaligned). This absence of interaction effect may be due to the substantial influence of perceived source credibility and political affiliation on judgment processes within these political decision-making contexts \cite{michael2021relationship, roozenbeek2022susceptibility, swire2017processing}. In our experiment, each claim was presented alongside the speaker's name to mirror the presentation of real-world social media posts, where both the message content and the speaker name are visible. However, this setup might have obscured the distinction between content and social explanations in the political domain, as participants may have processed the content explanations in a manner that integrated social aspects such as the speaker's political affiliation and perceived credibility. For instance, if a speaker known for their conservative affiliations presented a claim about tax reform, participants might have interpreted the content explanation through the lens of the speaker's political background, considering potential conservative biases or motives influencing the message. Indeed, our analysis of participants' responses revealed many instances where participants highlighting potential political biases, indicated by speakers' political affiliations and contexts, as critical in their decision-making process. This emphasis on socio-contextual information in the political domain, combined with our experimental design, might have obscured the distinct cognitive mechanisms we discussed would be activated by different explanation types, explaining the absence of a significant interaction effect found in this domain.}

Finally, we note that discrepancies in explanation effectiveness across domains were primarily observed in \textit{weighted} detection accuracy, but not in standard detection accuracy. This distinction implies that while the presentation order and type of explanations may not alter users' binary detection accuracy, they significantly affect the confidence with which these judgments are made. Given that subjective decision confidence can influence individuals' willingness to seek further information and engage in critical thinking \cite{desender2018subjective}, our findings about the impact on weighted detection accuracy carry important implications: the presentation of explanations—whether in terms of type, order, or alignment—can profoundly influence decision confidence and, consequently, subsequent user behavior.

\subsubsection{Ethical Considerations of Deploying AI-Generated Explanations in Misinformation Detection}
\rblue{While our findings suggest that AI-generated explanations can help users detect misinformation, deploying them in real-world settings introduces several ethical concerns. One such concern is over-reliance of AI, where users uncritically accept AI-generated explanations, assuming their accuracy without independent verification \cite{buccinca2021trust}. Although over-reliance is a well-documented phenomenon in XAI research \cite{buccinca2021trust}, our study found limited direct evidence supporting over-reliance on explanations. Instead, we observed that participants engaged with explanations in varied ways, including actively evaluating them, exhibiting confirmation bias, or selectively disregarding certain explanations. We believe this outcome may be due to our study design, which may have mitigated conditions that typically lead to over-reliance phenomenon. Specifically, we incorporated a screening test to ensure participants actively read and engaged with the explanations. Additionally, we did not provide the model's prediction for each claim, which may have encouraged participants to examine the explanations more carefully to infer the model's decision. While real-world applications may not always implement these safeguards, our findings highlight the potential of intervention strategies that promote active engagement with explanations as an effective approach to mitigating blind trust in AI-generated explanations \cite{buccinca2021trust, vasconcelos2023explanations, lee2023understanding, de2025cognitive}.} 

\rblue{Beyond over-reliance, bias in AI-generated explanations presents another critical challenge. LLMs inherit and reproduce biases related to gender, race, and political ideology from their training data \cite{weidinger2021ethical, bang2024measuring}, which can manifest in both content-based and social explanations. For instance, in content explanations, an AI model might disproportionately flag non-standard dialects—such as African American English (AAE)—as unreliable or misleading simply because they diverge from the model’s learned conventions of `trustworthy' language, which are often shaped by Western-centric linguistic norms \cite{hofmann2024ai}. Similarly, in social explanations, independent journalists or news outlets from underrepresented regions might be flagged as less credible due to a ``limited verification history,'' reinforcing existing disparities in information accessibility and trustworthiness \cite{techpolicyFairnessFactchecking}. Since our explanations were generated based on misinformation from PolitiFact, we did not observe these biases when reviewing both the content and social explanations. In other words, because PolitiFact primarily fact-checks English-language claims from United States \cite{gangopadhyay2024investigating} and relies on established media sources and institutional fact-checking standards\footnote{\href{https://www.politifact.com/article/2018/feb/12/principles-truth-o-meter-politifacts-methodology-i/\#How\%20we\%20choose\%20claims}{https://www.politifact.com/article/2018/feb/12/principles-truth-o-meter-politifacts-methodology}}, our explanation generation process did not interact with non-standard linguistic varieties or underrepresented news outlets. While evaluating potential biases in AI-generated explanations was beyond the scope of our work, future research could expand their datasets by incorporating fact-checks from non-Western sources, community-led initiatives, and linguistically diverse news outlets to examine how biases emerge in both content and social explanations across different linguistic, cultural, and geopolitical contexts.} 

\rblue{Finally, privacy concerns arise when AI-generated explanations, particularly social explanations, incorporate user-related data. For example, an explanation stating, ``This claim has been frequently shared by users who engage with politically biased or extremist content,'' may reveal sensitive user information—such as behavioral profiling (tracking repeated engagement with specific content), ideological inference (associating users with polarized or extremist viewpoints), and historical tracking (referencing past interactions with similar claims or sources)—that individuals neither expect nor wish to be disclosed. Such disclosures can also raise concerns about algorithmic surveillance and the unintended reputational harm of those implicated by such explanations (e.g., being unfairly labeled as politically biased or extremist, facing social stigma). In our study, privacy risks were minimal since we relied solely on publicly available socio-contextual cues and did not use private user data. However, in real-world applications—particularly on social media platforms, where vast amounts of user-generated data are readily accessible—deploying social explanations without strong safeguards could lead to potential privacy risks, algorithmic profiling, and possible misuse by platforms or third parties for targeted advertising or misinformation amplification.}

\rblue{Given the aforementioned ethical considerations, we argue that teams planning to deploy AI-generated explanations must systematically evaluate their potential impacts before real-world implementation. This evaluation may include conducting explanation audits to assess whether AI explanations reinforce biases or systematically disadvantage certain perspectives, performing extensive user testing to determine whether explanations improve decision-making or unintentionally reinforce over-reliance, and ensuring privacy safeguards when incorporating social explanations, particularly in contexts where user-generated data is involved. For research teams working on AI-generated explanations, we recommend further advancing ongoing efforts to mitigate over-reliance (e.g., refining explanation interfaces that encourage users to critically engage with AI-generated explanations), bias (e.g., improving debiasing frameworks to assess whether explanations systematically favor or disadvantage specific groups, dialects, or political perspectives), and privacy risks (e.g., improving privacy-preserving techniques such as differential privacy or content obfuscation to prevent AI-generated explanations from exposing sensitive user data). Additionally, further research is needed to examine how content and social explanations manifest in real-world settings, influence user behavior over time, and hold up in ecologically valid scenarios beyond controlled experiments. As we did in our study, we also encourage researchers to debrief participants that AI-generated explanations may be incorrect so that they do not leave with the impression that these explanations are inherently reliable and accurate.}

\subsection{Implications for Designing Effective AI Explanations}
Our study highlights the need of carefully designing AI explanations to enhance their effectiveness in misinformation detection. First, the consistent effectiveness of aligned explanations across two studies suggests a need to prioritize alignment between content and social explanations to improve both accuracy and user experience. \rblue{In our experiment using various prompts for explanation generation, when prompting the LLM to generate both content and social explanations \textit{together}, these explanations tended to be aligned and coherent in most cases, minimizing contradictions and reducing misalignment. However, contradictions between the content and social explanations still occurred in some instances, likely due to difficulty in reconciling the very different reasoning frameworks that underpin content-based and social explanations. In such cases, alignment-check mechanisms and misalignment correction strategies can be implemented to enforce consistency. One approach is a pre-generation alignment filter, where the AI system first assesses whether the content and social explanations contradict each other. If a misalignment is detected, the system can enforce a single, aligned explanation by prioritizing one perspective that aligns with the final decision and adjusting the other to fit within this framing. For example, if a content explanation deems a claim false due to sensational language, the system can reframe the social explanation to highlight that even credible sources occasionally share misinformation. However, while enforcing alignment is possible, overwriting one explanation to fit the final decision risks creating artificial coherence, leading to epistemic overreach, where the AI appears more confident than warranted.} 

Moreover, as discussed in Section \ref{sec:dis1}, misaligned explanations—despite potentially reinforcing confirmation biases—present a unique opportunity to foster deeper \textit{critical thinking}. Therefore, it is crucial to develop strategies that leverage the benefits of misaligned explanations while mitigating their risks. One promising approach is the proactive disclosure of conflicting information before presenting individual explanations, allowing users to recognize potential contradictions before forming their judgments. By forewarning users about inconsistencies, the system can encourage critical engagement and openness to alternative viewpoints \cite{turner2010lay}. \rblue{Furthermore, we observed that participants in the misaligned condition engaged with content-based and social explanations in distinct ways, often experiencing uncertainty and showing varied preferences and usage patterns when relying on explanations for decision-making. To support users who feel uncertain when faced with conflicting explanations—such as those who explicitly express confusion or take extended time to decide whether the claim is true or false—the system could provide dynamic prompts that guide them toward structured reasoning. For example, prompts like ``What specific aspects of each explanation make you more or less confident?'' or ``If you had to choose one explanation as more reliable, which would it be and why?'' could encourage deeper evaluation of misaligned explanations. Additionally, the system can present the model’s final prediction alongside LLM-generated confidence scores for each explanation \cite{lin2022teaching}, helping users evaluate the relative reliability of different perspectives and make more informed judgments \cite{rechkemmer2022confidence}. To accommodate individual differences in explanation preferences, the system could allow users to customize how explanations are presented—offering options such as side-by-side comparisons, where both content and social explanations are displayed together for direct evaluation, or sequential exploration, where users can choose which explanation to view first while being prompted to consider the alternative afterward.} \mt{Together, these strategies reflect a set of structured interaction mechanisms designed to support users across key stages of engagement with misaligned explanations---beginning with early-stage misalignment forewarning, followed by dynamic prompting during moments of uncertainty, and further supported by confidence visualization for evaluative reasoning and adaptive explanation presentation that responds to user behavior throughout the interaction. We propose to formalize these mechanisms as sequential, user-centered interaction steps under the Proactive User Engagement Paradigm (PUEP), which we introduce in the Section 6.3.}

Finally, our findings reveal variability in the effectiveness of explanations across different domains, emphasizing the need for adaptive strategies tailored to each context. For instance, the significant interaction effect between the explanation type and alignment was observed in the COVID-19 domain. This finding suggests that careful attention should be given to both the type of explanation presented first and how well subsequent explanations align with or complement the initial one, as this combined influence can be crucial for maximizing user understanding and detection accuracy. \rzre{Conversely, in the political domain, no interaction effect was observed, which we hypothesize may stem from participants’ emphasis on socio-contextual information when evaluating political claims. If this is the case, a more deliberate design of social explanations—such as integrating accuracy prompts within the explanation process (e.g., ``Think whether the explanation is accurate'') \cite{pennycook2021shifting}—could be beneficial.}

\subsection{The Theoretical Contributions to CSCW and XAI}
\rblue{Our findings provide a structured foundation for designing AI-generated explanations for misinformation detection, demonstrating that the way explanations are generated, aligned, and structured significantly impacts user experience and detection accuracy. Building on these findings, we propose the Multi-Stage Explanation Framework (MSEF), which organizes explanation generation and presentation in three key stages: (1) Independent explanation generation, (2) Alignment assessment and mediation, and (3) Dynamic explanation structuring. Specifically, \textit{Independent explanation generation} stage ensures that multiple perspectives (e.g., content-based and social explanations) are fully developed without oversimplification and premature convergence. The \textit{Alignment assessment and mediation} stage involves the system evaluating whether explanations are aligned and, if misaligned, determining how to present, adjust, or contextualize them to reduce uncertainty and guide the user’s critical evaluation. If misalignment between explanations is detected, for example, the system can provide a contextual explanation detailing potential reasons for divergence (e.g., differences in the aspects each explanation type focuses on), helping users critically evaluate these discrepancies and make informed judgments rather than feeling confused by the need to determine whether one explanation is entirely correct or incorrect. Finally, the \textit{Dynamic explanation structuring} stage adapts explanations based on the domains, user preferences, and different usage patterns. This adaptability not only allows users to tailor how explanations are presented-such as the customization options discussed earlier in Section 6.2-but also dynamically adjusts explanation formats based on users' real-time interaction patterns. For example, if a user consistently prefers one explanation over another without engaging with the alternative when both explanations are presented simultaneously, the system could switch to a sequential format to encourage them to consider both explanations before making a judgment. Beyond introducing this novel framework to the CSCW and XAI communities, our results highlight the need for new Proactive User Engagement Paradigm (PUEP) in these research areas to help users navigate misaligned explanations—especially given that misalignment occurred frequently (58.04\% in our study). Conducting an experiment based on our design considerations, previously discussed in Section 6.2—such as helping users follow a structured reasoning process when encountering conflicting explanations—could inform the operationalization of this new paradigm for CSCW, revealing how users interpret conflicting explanations, what factors encourage deeper analysis, and how explanation effectiveness varies depending on individual differences (e.g., cognitive styles, prior beliefs, expertise). Altogether, our study contributes to ongoing discussions in XAI and CSCW by shedding light on the nuanced ways explanation presentation shapes user engagement and decision-making. More broadly, these findings suggest new research directions for these communities to explore: (1) how users process misaligned explanations to better understand the underlying cognitive mechanisms; (2) how explanation effectiveness varies across domains, in order to identify domain-specific challenges and best practices; and (3) how to design AI-generated explanations that integrate content-based and social perspectives to better support critical reasoning in misinformation detection.}

\subsection{Future Work}
In this paper, we have taken an initial step towards exploring how different explanation types and their combinations can significantly affect misinformation detection accuracy and user experience. Based on the preliminary findings from this study, we plan to conduct more confirmatory studies in the future to validate our findings and assess their generalizability across diverse settings. Our research so far has been constrained to natural language explanations generated by GPT-4o in COVID-19 and Politics domains. It remains unclear whether the effectiveness of different explanation types (content, social, aligned, misaligned explanations) and their impact on user experience are consistent across different domains, such as entertainment, climate, \mt{or financial misinformation}. One potential outcome we anticipate when expanding our study to different domains is that while overall trends in explanation effectiveness may remain consistent, the specific ways explanations influence user decision-making could vary. More specifically, in our study, despite the distinct characteristics of COVID-19 and political misinformation \cite{muhammed2022disaster}, we found that aligned explanations were consistently effective across both domains, whereas misaligned explanations were generally less effective. However, the ordering effect of content-based and social-based explanations differed between the two, \mt{suggesting that the cognitive mechanisms activated by content- and social-based explanations may vary with the misinformation domain.} \mt{It therefore remains an open question whether these patterns hold in other misinformation domains and whether domain-specific factors (e.g., topic politicization, urgency) modulate explanation impact. For instance, in climate misinformation where probabilistic reasoning about future temperature trajectories or risk projects are central to public discourse \cite{cook2020deconstructing}, content-based explanations may play a more prominent role in shaping users' credibility judgments as users may expect analytic, data-driven argumentation in this context. As a result, explanation-order effects are likely to be weakened---similar to our findings in the political misinformation---unless the climate issues become highly politicized (e.g., when mitigation policies are framed in partisan or identity-based terms). In entertainment misinformation, by contrast, social explanations may exert greater influence, given that such content often circulates through influencer networks and fan communities \cite{lee2022community}, where group identity and emotional resonance can outweigh the persuasiveness of content explanations. In the context of financial misinformation, the strength of explanation order effects likely depends on the interaction between users' motivational states and cognitive resources---specifically, the tension between the desire for high gains and the motivation for accuracy \cite{rangapur2023investigating}. When urgency-driven narratives  (e.g., ``limited-time offers'', ``secret investment tips'') heightens users' aspiration for quick rewards, they may become more susceptible to initial social explanations, which can anchor user expectations and reduce openness to subsequent content-based explanations offering analytical or statistical caution. However, when users perceive substantial financial risk, accuracy motivations may become more salient, potentially weakening, or even reversing, the influence of the social-first explanation order.} \mt{W}e believe that our explanation generation approach \mt{offers a flexible framework for testing these hypotheses across diverse misinformation domains.} Specifically, our explanation generation pipeline (see Appendix \mt{B}) is domain-agnostic and does not rely on rigid pre-defined mappings (e.g., associating emotional language with misinformation) but instead leverages GPT’s pre-trained knowledge to generate explanations. 

\rblue{In addition, our explanation design aimed to ensure rigor by isolating the effects of different explanation types without the influence of a final verdict. However, this approach comes with limitations in external validity, as real-world misinformation detection systems typically present both a verdict and accompanying explanations. This discrepancy between controlled experimental conditions and real-world systems raises a critical question: How does the presence of a final verdict shape user engagement with explanations, and how do these effects compare to our study’s findings? A natural next step for future work is to conduct a similar experiment that presents a model's final verdict alongside explanations to examine its impact on user engagement and decision-making. One plausible outcome is that verdicts may reduce users’ motivation to engage critically with the explanation, leading to cognitive shortcuts and overreliance on AI outputs \cite{gajos2022people}. This concern is especially relevant in conditions where explanations are misaligned, as users may defer to the verdict as an authoritative resolution rather than attempt to resolve inconsistencies themselves. Thus, another crucial research direction is exploring how to mitigate overreliance on verdicts when explanations are misaligned. Several open questions emerge from this line of work. For instance, would greater transparency into how a verdict is reached encourage deeper engagement with misaligned explanations, helping users critically assess inconsistencies? Or, would it instead reinforce trust in the verdict, making the model appear even more authoritative and further discouraging scrutiny despite conflicting explanations? Moreover, do users interpret a final verdict as the result of aggregating two conflicting explanations? If so, how does the way this aggregation process is communicated—for example, through structured comparisons or high-level summaries—affect user trust, critical evaluation, and engagement with both the final verdict and the explanations? We encourage future research to explore these important and interesting questions.}

Finally, we observed different effectiveness of explanations across domains and hypothesized that this result may be due to content and social explanations triggering different cognitive processing styles, and also depending on what users deem important within each domain when making decisions. We recommend that researchers conduct experiments to further validate this hypothesis, such as using eye-tracking and response time measurements to assess how users cognitively process content versus social explanations across different contexts. Ultimately, our goal is to gain deeper understanding of how different explanation types and presentation strategies impact people's ability to detect misinformation, which can help inform the design of a more effective AI explanations of misinformation. 

%% file: 6conclusion.tex
\section{Conclusion}
\label{sec:conclusion}
In this paper, we examine the effects of various AI explanations on users' ability to detect misinformation and their overall user experience. In particular, we design and compare content explanations, social explanations, and combined explanations, where content and social explanations either align or misalign with each other. Our results across two \rzre{studies}
reveal that AI explanations can effectively assist users in detecting misinformation in many cases. We also discovered that the 
presentation order and the alignment between content and social explanations play significant roles in the effectiveness of AI explanations, with variations across domains. Based on these findings, we position that AI explanations should be carefully designed to consider 
\rzre{the domain topic, presentation order,} and the alignment between content and social explanations to enhance their impact on misinformation detection and user experience. Our study provides valuable insights for designing more effective misinformation explanations and opens the door for future research in the realm of human-centered explainable AI (XAI).


%% file: 7appendix.tex
\section*{Appendix}
\section{Rationale for Separately Generating Content- and Social-Based Explanations}
\mt{While generating content- and social-based explanations separately may seem artificial, this design choice was necessary to systematically study misaligned explanations. Examining misaligned explanations is important because conflicting credibility signals frequently occur in real-world misinformation evaluation (e.g., \cite{baum2021emotional, avram2020exposure}). For example, on social media platforms, users may encounter claims that seem factually inaccurate based on content-based cues—such as emotional language, exaggerated phrases, or informal language-yet originate from sources perceived as credible. By intentionally separating content- and social-based explanations, we can examine how users perceive these conflicts and how misalignment influences their decision-making and overall experience—an area that has been underexplored in prior work. Similarly, misinformation detection models also struggle to align content- and socio-contextual cues, often resolving inconsistencies through weighting mechanisms that aggregate conflicting signals into a single prediction and explanation \cite{guo2023two, prakash2023fake}. However, this aggregation process obscures underlying inconsistencies rather than making them explicit to users, preventing a deeper understanding of how conflicting signals shape misinformation assessments. Aggregating conflicting signals into a single output also results in a missed opportunity to design effective misinformation explanation systems that explicitly reveal these inconsistencies and foster critical thinking, especially if highlighting conflicting credibility cues through misaligned explanations proves to be beneficial. Furthermore, the prevalence of misaligned cues or explanations in real-world misinformation remains largely unknown, as existing approaches prioritize aggregation over examination of inconsistencies. Our work addresses this gap by finding that 58.04\% of the generated explanations were categorized as misaligned, while 41.96\% were aligned, highlighting the possibility of a significant presence of misaligned explanations in the real world.}

\section{Prompt Design for Explanation Generation}

In our study, we developed a prompting template aimed at effectively generating content and social explanations. This template was refined iteratively to enhance context-relevance, specificity, and consistency within and across the generated explanations. We assigned the role of ``AI Explanation Expert'' to GPT-4o for both content and social analysis. For guiding these analyses, we provided detailed instructions focusing on specific cues: content cues for content analysis and socio-contextual cues for social analysis, to ensure the context relevance and specificity of the explanations. We also aimed for consistency in format across all explanations by defining uniform formats and constraints. 
We used the GPT-4o-2024-08-06 from OpenAI's API, setting the temperature parameter to t=0.0 and limiting the maximum token count to 150 for conciseness, 2 to 3 sentence responses for each explanation type. 

For example, the prompt used for generating content explanation is as follows:

\begin{quote}
    \tt\small You are an AI explanation expert specialized in content analysis, focusing on linguistic features of claims. Your task is to analyze the syntax, semantics, and structure to evaluate the veracity of the claim. \\
    Claim: \{\text{CLAIM}\} \\
    Based on your analysis `exclusively' related to linguistic cues typically found in misinformation, how would you categorize the above claim? A: True, B: False\\
    Format: Decision (either True or False), Explanation. \\
    Constraints: Your explanation should focus on the specific linguistic and structural aspects of the claim,  guiding users to understand how these features inform the veracity of the claim.\\
\end{quote}

Similarly, the prompt used for generating social explanations is as follows: 

\begin{quote}
    \tt\small You are an AI explanation expert specialized in social analysis, focusing on speaker's attributes and the context of the claim. Your task is to analyze the socio-contextual cues provided to evaluate the veracity of the claim.  \\
    Claim: \{\text{CLAIM}\} \\
    Socio-contextual cues: \{\text{SPEAKER}\}, \{\text{JOB TITLE}\}, \{\text{PARTY AFFILIATION}\}, \{\text{CREDIT HISTORY}\}, \{\text{CONTEXT}\} \\
    Based on your analysis `exclusively' related to socio-contextual cues typically found in misinformation, how would you categorize the above claim? A: True, B: False\\
    Format: Decision (either True or False), Explanation. \\
    Constraints: Your explanation should focus on the specific socio-contextual aspects of the claim, guiding users to understand how these features inform the veracity of the claim.\\
\end{quote}

However, it is important to note that although our template was the best at producing relevant and specific explanations compared to other prompting methods we experimented with, several limitations were observed. First, there were instances where the GPT-4o model deviated from our instructions to focus exclusively on the cues provided, such as referencing historical records or specific actions of the speaker, possibly due to the model's generative nature. Sometimes, GPT-4o provided ``Unknown'' verdicts despite the instruction to choose between either True or False. We also observed some vague and uninformative explanations generated, such as ``The analysis the claim indicate that it is inaccurate.'' 
In all these instances, we manually revised the explanations to ensure their clarity and accuracy. 

\rblue{In addition, we emphasize that our explanation generation approach can be adapted for use in other domains besides Politics and Covid-19. Researchers aiming to extend our findings can apply the method by providing the claim (to generate the content-based explanation) and relevant socio-contextual cues (to generate the social-based explanation) based on their domain of interest. However, certain modifications may be necessary to ensure that the generated explanations align with domain-specific misinformation strategies. For example, in the context of climate misinformation, researchers may consider refining the explanation prompts to better detect misleading statistical projections—such as selectively reported temperature trends that omit long-term climate patterns—and incorporating socio-contextual cues that reflect expert scientific consensus rather than relying on generic credibility indicators [14].} 
 
\section{Claims and Explanations Used in the Experiment}
Below are examples of claims used in our experiments, along with their respective content and social explanations.\footnote{The examples mentioned, including references to public figures in the explanations, are solely for illustrative purposes and not meant to be offensive or defamatory. These explanations are informed by information from PolitiFact, a well-regarded fact-checking website that is accessible online. We also disclose at the end of the experiment that the explanations are not guaranteed to be accurate as they are generated by an AI model (GPT), and all sources referenced in our research have been shared for transparency.} These examples represent each category across two domains (COVID-19, Politics) and two ground truth labels (True, False).

\subsection{Domain: COVID-19 \& Ground Truth Label: True}
\begin{itemize}

    \item \textbf{Claim}: No city in the state can quarantine itself without state approval. 
    \item \textbf{Content Explanation}: The claim's use of the absolute term ``No'' at the beginning of his statement establishes a definitive and blanket policy, which can suggest an oversimplification. The use of passive voice in "can quarantine itself without state approval" reduces the clarity on who or what entity is responsible for the approval process, potentially leading to ambiguity.    
    \item \textbf{Social Explanation}: As the Governor of New York, Andrew Cuomo would be expected to have a thorough understanding of state policies and regulations, especially those related to public health and safety during a pandemic. The context of a news conference, a setting where public statements are scrutinized and widely disseminated, would typically motivate a public official to ensure the accuracy of their statements.
\end{itemize}

\subsection{Domain: COVID-19 \& Ground Truth Label: False}
\begin{itemize}
    \item \textbf{Claim}: All these athletes are dropping dead on the field after receiving the COVID-19 vaccination.
    \item \textbf{Content Explanation}: This claim uses the dramatic term ``dropping dead,'' which introduces a sensational and hyperbolic tone. This kind of language is often associated with exaggeration rather than a factual report. The absence of specific proper nouns or detailed comparative data to substantiate the claim makes it less credible. The overall emotional and subjective nature of the statement, designed to evoke fear or concern, further detracts from its objectivity and aligns more with misinformation or unverified claims.
    \item \textbf{Social Explanation}: As a Republican U.S. Senator from Wisconsin, Ron Johnson's political affiliation might suggest potential biases, particularly in the polarized context of public health and COVID-19 responses. The setting of a radio interview often allows for more informal conversation and personal opinions, which may not always be grounded in rigorous evidence. This context, coupled with Johnson's political background, may indicate that the claim could be driven more by personal or political viewpoints rather than substantiated scientific data. 
\end{itemize}

\subsection{Domain: Politics \& Ground Truth Label: True}
\begin{itemize}
    \item \textbf{Claim}: At Bain Capital, we helped start an early childhood learning company called Bright Horizons that First Lady Michelle Obama rightly praised. 
    \item \textbf{Content Explanation}: The phrase ``helped start an early childhood learning company called Bright Horizons'' is straightforward and lacks ambiguity, which typically characterizes truthful claims. The simple subject-verb-object format and absence of linguistic indicators like exaggeration or vagueness further support the claim's veracity. 
    \item \textbf{Social Explanation}: As a former Republican governor of Massachusetts and his long-term association with Bain Capital, Romney has a credible basis for knowledge about the company's initiatives. Additionally, the setting of the Republican National Convention, known for its heightened scrutiny, implies a commitment to accuracy. 
\end{itemize}

\subsection{Domain: Politics \& Ground Truth Label: False}
\begin{itemize}
    \item \textbf{Claim}: Barack Obama's plan calls for ``mandates and fines for small businesses.''
    \item \textbf{Content Explanation}: The claim that the plan calls for ``mandates and fines for small businesses'' lacks specific detail and context, which is crucial in understanding policy implications. The use of broad terms like ``mandates'' and ``fines'' without further elaboration can be misleading, suggesting a more burdensome impact on small businesses than what might be the case. 
    \item \textbf{Social Explanation}: The context the claim was made, a political debate, is known for its strategic rhetoric and partisan viewpoints, which can sometimes lead to exaggerated or distorted statements. Also, McCain's varied credibility history suggests a tendency towards statements that are not consistently factual, further casting doubt on the veracity of his claim. 
\end{itemize}

\section{Further Analysis on Misaligned Condition}

\rblue{To examine whether misaligned explanations contributed to confirmation bias, we conducted two analyses: First, we compared participants' initial decisions (made before viewing the explanation) with their revised decisions (made after viewing the explanation). We defined the ``Average Change in Decision'' as the proportion of change in participant's decisions after viewing the explanations, calculated by the average number of decision changes divided by the total number of claims participants decided on. A closer value to 0 indicates that participants' final decisions closely matched their initial ones, while larger values indicate greater shifts away from their initial decisions. Second, we examined participants’ confidence shifts before and after viewing the explanation, focusing specifically on claims where their decision remained unchanged.\footnote{\rblue{We focused on claims where participants' decisions did not change because confidence shifts in these cases reveal whether explanations reinforce initial judgments rather than prompting reconsideration. In contrast, for claims where decisions changed, confidence shifts could be influenced by the act of changing one's decision itself, making it difficult to isolate the effect of the explanation on confidence alone.}} We defined ``Average Change in Confidence'' as the mean difference in a participant’s confidence ratings for the claims where their decisions remain unchanged, calculated by averaging the change in confidence for each of these individual claim. Positive values indicate increased confidence after the explanation, while negative values indicate decreased confidence. A value closer to 0 suggests minimal change in confidence, whereas larger absolute values reflect stronger belief shifts. For these two analyses, we compared decision changes and confidence shifts in the misaligned condition with those in the aligned, content, and social explanation conditions to assess the significance of the observed findings.} 

\begin{figure}[htb!]
    \includegraphics[width=0.9\columnwidth]{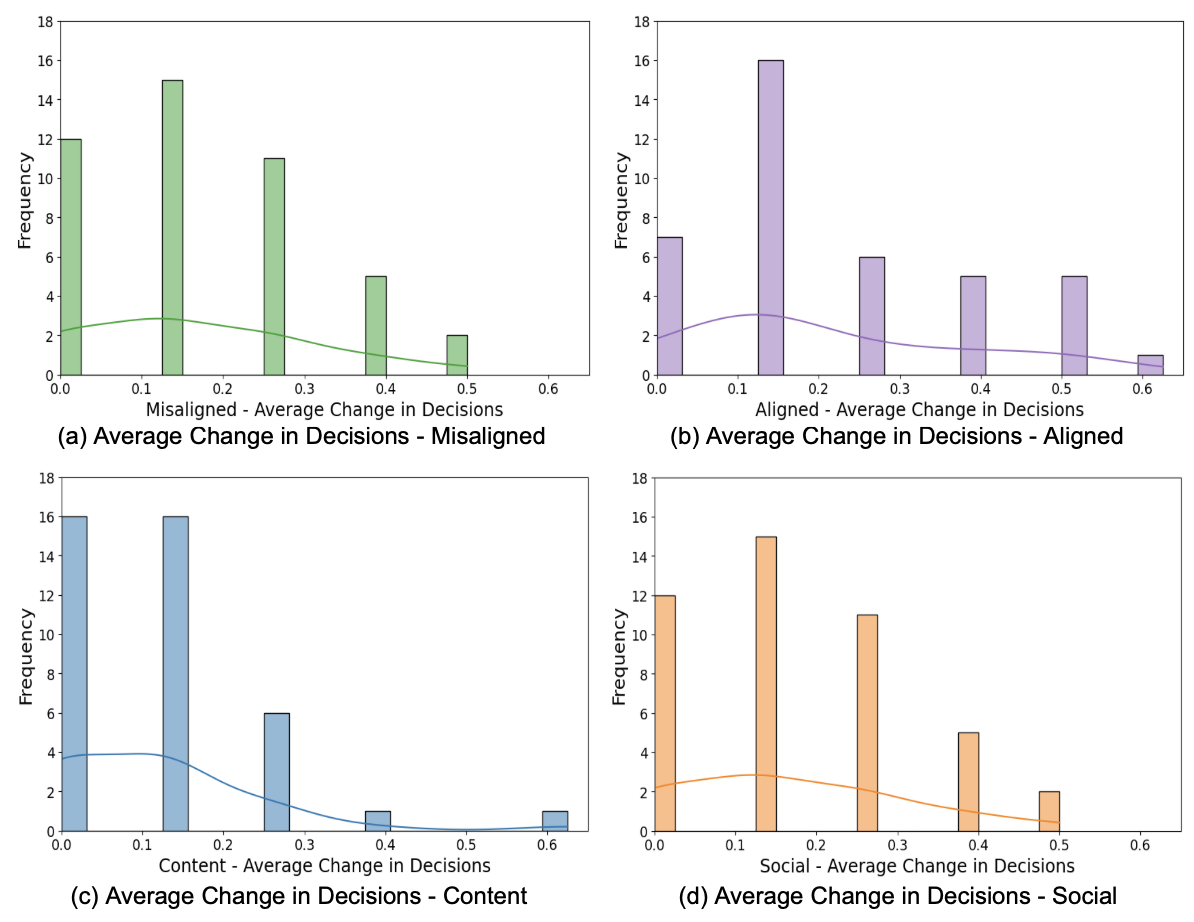 }
    \caption{Decision changes across all conditions: Effects of presenting misaligned (3a), aligned (3b), content (3c), and social (3d) explanations. ``Average Change in Decision'' is the proportion of change in participant's decisions after viewing the explanations.}
    \label{fig:abcd}
\end{figure}

\begin{figure}[htb!]
    \includegraphics[width=0.9\columnwidth]{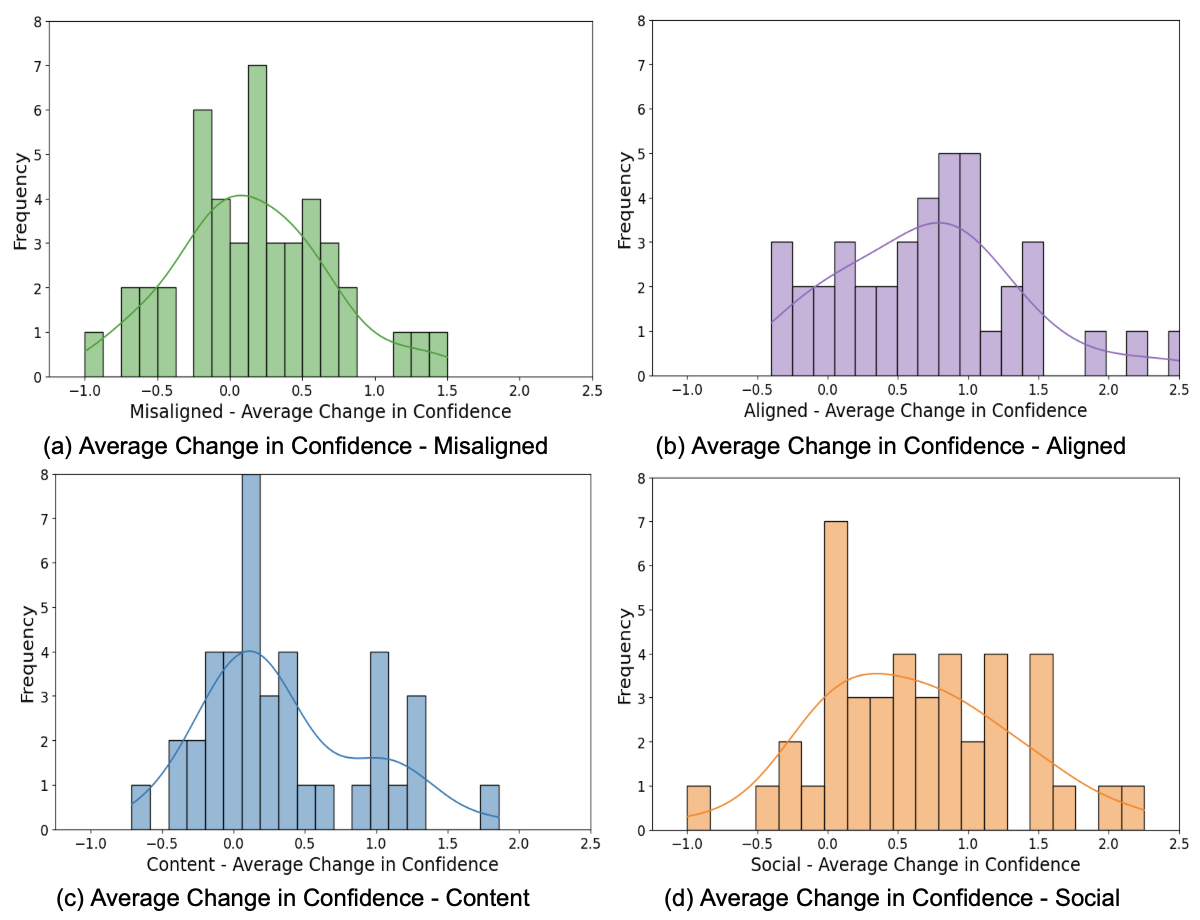 }
    \caption{Confidence changes across all conditions: Effects of presenting misaligned (4a), aligned (4b), content (4c), and social (4d) explanations. ``Average Change in Confidence'' is the average mean difference of change in participant's decisions after viewing the explanations. In this analysis, we focused only on claims where participants did not change their initial decisions after viewing the explanation.}
    \label{fig:gpgp}
\end{figure}
    
\rblue{Our comprehensive analysis of decision changes and confidence changes across different conditions is presented in Figure \ref{fig:abcd} and Figure \ref{fig:gpgp}, respectively. As hypothesized, Figure \ref{fig:abcd}(a) shows that participants generally adhered to their initial choice when presented with misaligned explanations. This suggests that, under misaligned conditions, participants tend to follow their original choices rather than reconsidering or changing their decisions. However, as seen in Figure \ref{fig:abcd}(b-d), participants across all conditions exhibited a general reluctance to change their decisions, indicating that this tendency was \textit{not} unique to the misaligned condition. Additionally, no statistically significant differences were found across the conditions in terms of decision change rates. When analyzing confidence shifts for cases where participants chose not to change their decisions in the misaligned explanation condition, we found that while confidence generally increased for these unchanged decisions (Figure \ref{fig:gpgp}(a)), this increase was small and did not provide strong evidence of a meaningful effect, nor was it significantly greater than in other conditions (Figure \ref{fig:gpgp}(b-d)). In addition, ANOVA results (F = 3.96, p = 0.014), followed by Tukey’s HSD post-hoc test, revealed that confidence increased significantly more in the aligned condition compared to both the content (p = 0.040) and misaligned conditions (p = 0.022), and no other pairwise comparisons being significant. Taken together, the findings from the two analyses suggest that while misaligned explanations may play a role in reinforcing confirmation bias, this effect does not appear to be exclusive to misaligned explanations and could instead reflect a broader pattern observed across different explanation types. Thus, additional interpretations is needed to explain the discrepancy between objective decision-making accuracy and subjective perceived usefulness in the misaligned explanation condition. Moreover, the observed stability in decisions should not be interpreted as evidence that the explanations were ineffective. Rather, it suggests that the effectiveness of explanations may manifest in more nuanced ways—such as shaping how participants interpret and make sense of the claim by influencing their information processing and fostering critical thinking—rather than directly prompting decision reversals or confidence shifts. This interpretation is further supported by our qualitative findings in Study 2, where we found that participants actively engaged with the explanations in varied and reflective ways.}

\begin{figure}[htb!]
    \includegraphics[width=0.85\columnwidth]{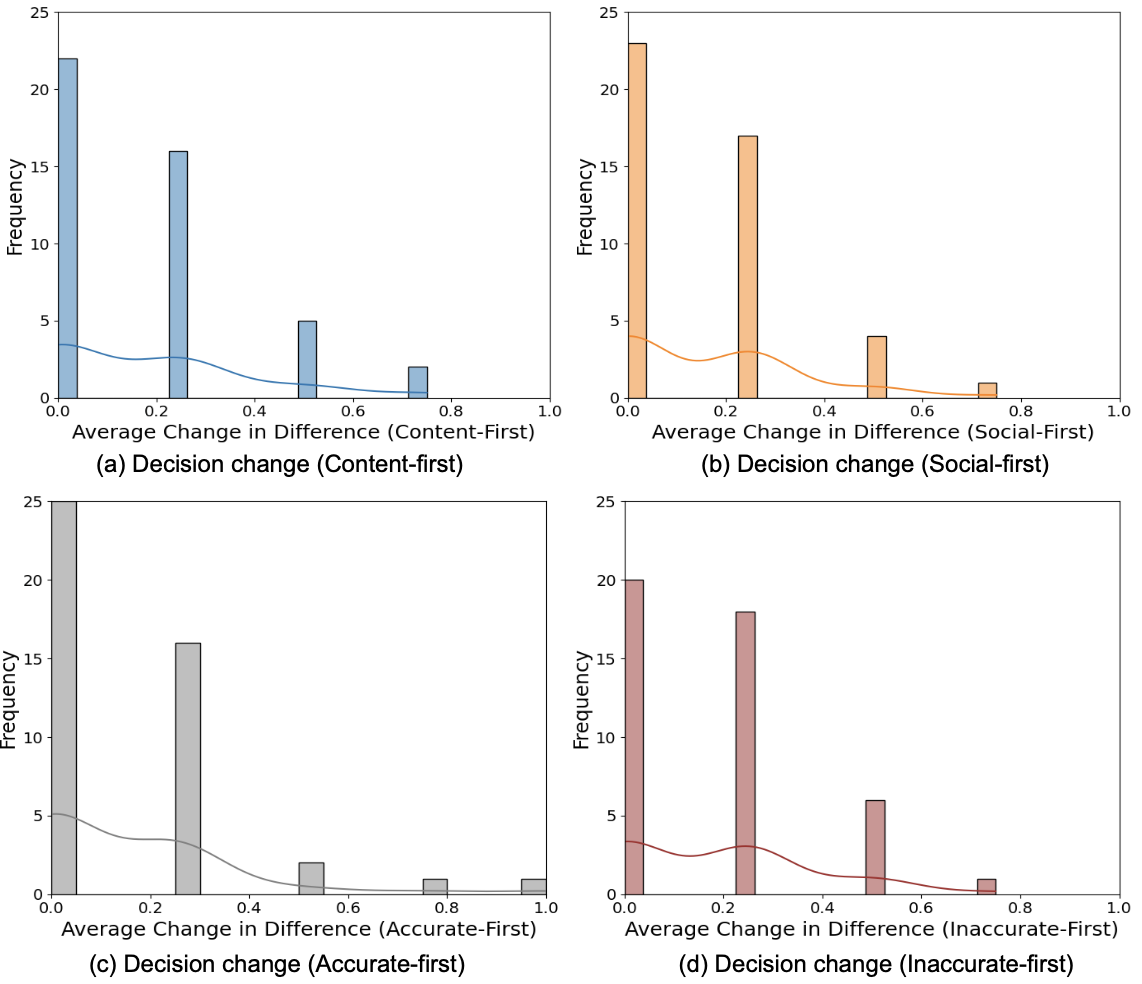 }
    \caption{Decision changes under misaligned conditions: Effects of presenting content (5a) and social (5b) explanations first;  Effects of presenting accurate (5c) and inaccurate (5d) explanations first.}
    \label{fig:aaaa}
\end{figure}

\rblue{In addition, our finding that explanations—particularly misaligned explanations—do not generally lead to changes in participants' decisions or significantly impact their confidence naturally raises an important question: How can we present misaligned explanations more effectively to help users make better judgments? One potential factor to consider is the order in which explanations are presented—could the sequence influence how likely participants are to revise their decisions?} To answer this question, we investigated whether the sequence in which explanations are presented influences participants' likelihood of changing their decisions. To this end, we examined two key dimensions of order (see Figure \ref{fig:aaaa} (a-d)): the type of explanation presented first (whether content or social explanation is presented first) and the accuracy of the first explanation (whether accurate or inaccurate explanation is presented first).\footnote{Our experimental design systematically counterbalanced both the order and accuracy of these explanations, enabling us to assess how these factors influenced participants' decision-making. Although both explanations were displayed simultaneously to users, they were arranged in top and bottom positions. For our analysis, we defined the `first' explanation as the one displayed at the top and the `second' explanation as the one displayed at the bottom, following the typical top-to-bottom reading order \cite{gobel2015up}.} However, our analysis suggests that neither the sequence in which explanations were presented (content-first vs. social-first) (Figure \ref{fig:aaaa} (a,b)) nor the accuracy of the first explanation (accurate-first vs. inaccurate-first) (Figure \ref{fig:aaaa} (c,d)) had a substantial effect on participants' decision changes. That is, we found no statistically significant differences between either the order of explanation type or the order of explanation accuracy, suggesting that the order in which explanations are provided does not significantly influence how participants revise their decisions under misaligned conditions. \rblue{Similarly, in the aligned explanation condition, presentation order—whether content or social, accurate or inaccurate—also had no meaningful impact on participants' decision revisions, indicating that explanation order does not play a significant role in influencing decision changes, regardless of alignment.} 

Together, these findings provide a nuanced perspective on misaligned explanations: participants tend to adhere to their initial decisions regardless of the type or accuracy of the first explanation they encounter. One possible reason for this could be the way explanations were presented; the current setting, where explanations were displayed simultaneously with the first explanation positioned at the top, might not have been strong enough to induce the priming effect typically associated with the first explanation. Additionally, asking participants to assess whether a claim was true or false before viewing the explanations may have primed them to reinforce their initial stance. Therefore, in Study 2, we explore whether omitting this pre-decision step and implementing a more explicit priming effect can reveal different patterns in responses to misaligned explanations.

\section{Study 2 - Standard Detection Accuracy Results} In both the COVID-19 and politics domain, a mixed ANOVA revealed no significant main effect of explanation type on detection accuracy in both domains (COVID-19: F(1,79) = 0.009, \( p = 0.924 \); Politics: F(1,70) = 0.011, \( p = 0.915 \)). However, a significant main effect of alignment was observed in both domains (COVID-19: F(1,79) = 58.831, \( p < 0.001 \), \(\eta_p^2\) = 0.427; Politics: F(1,70) = 100.859, \( p < 0.001 \), \(\eta_p^2\) = 0.590). No interaction effect between the explanation type and alignment was found in both domains (COVID-19: F(1,79) = 0.772, \( p = 0.382 \); Politics: F(1,70) = 1.695, \( p = 0.199 \)).

%% file: main.bbl

\begin{thebibliography}{96}


\ifx \showCODEN    \undefined \def \showCODEN     #1{\unskip}     \fi
\ifx \showDOI      \undefined \def \showDOI       #1{#1}\fi
\ifx \showISBNx    \undefined \def \showISBNx     #1{\unskip}     \fi
\ifx \showISBNxiii \undefined \def \showISBNxiii  #1{\unskip}     \fi
\ifx \showISSN     \undefined \def \showISSN      #1{\unskip}     \fi
\ifx \showLCCN     \undefined \def \showLCCN      #1{\unskip}     \fi
\ifx \shownote     \undefined \def \shownote      #1{#1}          \fi
\ifx \showarticletitle \undefined \def \showarticletitle #1{#1}   \fi
\ifx \showURL      \undefined \def \showURL       {\relax}        \fi
\providecommand\bibfield[2]{#2}
\providecommand\bibinfo[2]{#2}
\providecommand\natexlab[1]{#1}
\providecommand\showeprint[2][]{arXiv:#2}

\bibitem[Abigail Adu-Daako(2024)]%
        {techpolicyFairnessFactchecking}
\bibfield{author}{\bibinfo{person}{Aishwarya~Vardhana Abigail Adu-Daako}.} \bibinfo{year}{2024}\natexlab{}.
\newblock \bibinfo{title}{{T}he {F}airness of {F}act-checking and {I}ts {I}mpact on {S}ocial {M}edia | {T}ech{P}olicy.{P}ress --- techpolicy.press}.
\newblock \bibinfo{howpublished}{\url{https://www.techpolicy.press/the-fairness-of-the-factchecking-and-its-impact-on-social-media/}}.
\newblock


\bibitem[Avram et~al\mbox{.}(2020)]%
        {avram2020exposure}
\bibfield{author}{\bibinfo{person}{Mihai Avram}, \bibinfo{person}{Nicholas Micallef}, \bibinfo{person}{Sameer Patil}, {and} \bibinfo{person}{Filippo Menczer}.} \bibinfo{year}{2020}\natexlab{}.
\newblock \showarticletitle{Exposure to social engagement metrics increases vulnerability to misinformation}.
\newblock \bibinfo{journal}{\emph{arXiv preprint arXiv:2005.04682}} (\bibinfo{year}{2020}).
\newblock


\bibitem[Balog et~al\mbox{.}(2023)]%
        {balog2023measuring}
\bibfield{author}{\bibinfo{person}{Krisztian Balog}, \bibinfo{person}{Filip Radlinski}, {and} \bibinfo{person}{Andrey Petrov}.} \bibinfo{year}{2023}\natexlab{}.
\newblock \showarticletitle{Measuring the Impact of Explanation Bias: A Study of Natural Language Justifications for Recommender Systems}. In \bibinfo{booktitle}{\emph{Extended Abstracts of the 2023 CHI Conference on Human Factors in Computing Systems}}. \bibinfo{pages}{1--8}.
\newblock


\bibitem[Bang et~al\mbox{.}(2024)]%
        {bang2024measuring}
\bibfield{author}{\bibinfo{person}{Yejin Bang}, \bibinfo{person}{Delong Chen}, \bibinfo{person}{Nayeon Lee}, {and} \bibinfo{person}{Pascale Fung}.} \bibinfo{year}{2024}\natexlab{}.
\newblock \showarticletitle{Measuring Political Bias in Large Language Models: What Is Said and How It Is Said}.
\newblock \bibinfo{journal}{\emph{arXiv preprint arXiv:2403.18932}} (\bibinfo{year}{2024}).
\newblock


\bibitem[Barman and Colan(2023)]%
        {barman2023does}
\bibfield{author}{\bibinfo{person}{Dipto Barman} {and} \bibinfo{person}{Owen Colan}.} \bibinfo{year}{2023}\natexlab{}.
\newblock \showarticletitle{Does Explanation Matter? An Exploratory Study on the Effects of Covid--19 Misinformation Warning Flags on Social Media}. In \bibinfo{booktitle}{\emph{2023 10th International Conference on Behavioural and Social Computing (BESC)}}. IEEE, \bibinfo{pages}{1--7}.
\newblock


\bibitem[Baum and Abdel~Rahman(2021)]%
        {baum2021emotional}
\bibfield{author}{\bibinfo{person}{Julia Baum} {and} \bibinfo{person}{Rasha Abdel~Rahman}.} \bibinfo{year}{2021}\natexlab{}.
\newblock \showarticletitle{Emotional news affects social judgments independent of perceived media credibility}.
\newblock \bibinfo{journal}{\emph{Social Cognitive and Affective Neuroscience}} \bibinfo{volume}{16}, \bibinfo{number}{3} (\bibinfo{year}{2021}), \bibinfo{pages}{280--291}.
\newblock


\bibitem[Bo et~al\mbox{.}(2023)]%
        {bo2023will}
\bibfield{author}{\bibinfo{person}{Hongbo Bo}, \bibinfo{person}{Yiwen Wu}, \bibinfo{person}{Zinuo You}, \bibinfo{person}{Ryan McConville}, \bibinfo{person}{Jun Hong}, {and} \bibinfo{person}{Weiru Liu}.} \bibinfo{year}{2023}\natexlab{}.
\newblock \showarticletitle{What Will Make Misinformation Spread: An XAI Perspective}. In \bibinfo{booktitle}{\emph{World Conference on Explainable Artificial Intelligence}}. Springer, \bibinfo{pages}{321--337}.
\newblock


\bibitem[Braun and Clarke(2012)]%
        {braun2012thematic}
\bibfield{author}{\bibinfo{person}{Virginia Braun} {and} \bibinfo{person}{Victoria Clarke}.} \bibinfo{year}{2012}\natexlab{}.
\newblock \bibinfo{booktitle}{\emph{Thematic analysis.}}
\newblock \bibinfo{publisher}{American Psychological Association}.
\newblock


\bibitem[Bu{\c{c}}inca et~al\mbox{.}(2021)]%
        {buccinca2021trust}
\bibfield{author}{\bibinfo{person}{Zana Bu{\c{c}}inca}, \bibinfo{person}{Maja~Barbara Malaya}, {and} \bibinfo{person}{Krzysztof~Z Gajos}.} \bibinfo{year}{2021}\natexlab{}.
\newblock \showarticletitle{To trust or to think: cognitive forcing functions can reduce overreliance on AI in AI-assisted decision-making}.
\newblock \bibinfo{journal}{\emph{Proceedings of the ACM on Human-computer Interaction}} \bibinfo{volume}{5}, \bibinfo{number}{CSCW1} (\bibinfo{year}{2021}), \bibinfo{pages}{1--21}.
\newblock


\bibitem[Carrasco-Farr{\'e}(2022)]%
        {carrasco2022fingerprints}
\bibfield{author}{\bibinfo{person}{Carlos Carrasco-Farr{\'e}}.} \bibinfo{year}{2022}\natexlab{}.
\newblock \showarticletitle{The fingerprints of misinformation: how deceptive content differs from reliable sources in terms of cognitive effort and appeal to emotions}.
\newblock \bibinfo{journal}{\emph{Humanities and Social Sciences Communications}} \bibinfo{volume}{9}, \bibinfo{number}{1} (\bibinfo{year}{2022}), \bibinfo{pages}{1--18}.
\newblock


\bibitem[Chaiken and Maheswaran(1994)]%
        {chaiken1994heuristic}
\bibfield{author}{\bibinfo{person}{Shelly Chaiken} {and} \bibinfo{person}{Durairaj Maheswaran}.} \bibinfo{year}{1994}\natexlab{}.
\newblock \showarticletitle{Heuristic processing can bias systematic processing: effects of source credibility, argument ambiguity, and task importance on attitude judgment.}
\newblock \bibinfo{journal}{\emph{Journal of personality and social psychology}} \bibinfo{volume}{66}, \bibinfo{number}{3} (\bibinfo{year}{1994}), \bibinfo{pages}{460}.
\newblock


\bibitem[Cheng et~al\mbox{.}(2019)]%
        {cheng2019explaining}
\bibfield{author}{\bibinfo{person}{Hao-Fei Cheng}, \bibinfo{person}{Ruotong Wang}, \bibinfo{person}{Zheng Zhang}, \bibinfo{person}{Fiona O'connell}, \bibinfo{person}{Terrance Gray}, \bibinfo{person}{F~Maxwell Harper}, {and} \bibinfo{person}{Haiyi Zhu}.} \bibinfo{year}{2019}\natexlab{}.
\newblock \showarticletitle{Explaining decision-making algorithms through UI: Strategies to help non-expert stakeholders}. In \bibinfo{booktitle}{\emph{Proceedings of the 2019 chi conference on human factors in computing systems}}. \bibinfo{pages}{1--12}.
\newblock


\bibitem[Choudhary and Arora(2021)]%
        {choudhary2021linguistic}
\bibfield{author}{\bibinfo{person}{Anshika Choudhary} {and} \bibinfo{person}{Anuja Arora}.} \bibinfo{year}{2021}\natexlab{}.
\newblock \showarticletitle{Linguistic feature based learning model for fake news detection and classification}.
\newblock \bibinfo{journal}{\emph{Expert Systems with Applications}}  \bibinfo{volume}{169} (\bibinfo{year}{2021}), \bibinfo{pages}{114171}.
\newblock


\bibitem[Cook(2020)]%
        {cook2020deconstructing}
\bibfield{author}{\bibinfo{person}{John Cook}.} \bibinfo{year}{2020}\natexlab{}.
\newblock \showarticletitle{Deconstructing climate science denial}.
\newblock \bibinfo{journal}{\emph{Research handbook on communicating climate change}} (\bibinfo{year}{2020}), \bibinfo{pages}{62--78}.
\newblock


\bibitem[Cui et~al\mbox{.}(2022)]%
        {cui2022meta}
\bibfield{author}{\bibinfo{person}{Jian Cui}, \bibinfo{person}{Kwanwoo Kim}, \bibinfo{person}{Seung~Ho Na}, {and} \bibinfo{person}{Seungwon Shin}.} \bibinfo{year}{2022}\natexlab{}.
\newblock \showarticletitle{Meta-path-based fake news detection leveraging multi-level social context information}. In \bibinfo{booktitle}{\emph{Proceedings of the 31st ACM International Conference on Information \& Knowledge Management}}. \bibinfo{pages}{325--334}.
\newblock


\bibitem[Danry et~al\mbox{.}(2023)]%
        {danry2023don}
\bibfield{author}{\bibinfo{person}{Valdemar Danry}, \bibinfo{person}{Pat Pataranutaporn}, \bibinfo{person}{Yaoli Mao}, {and} \bibinfo{person}{Pattie Maes}.} \bibinfo{year}{2023}\natexlab{}.
\newblock \showarticletitle{Don’t Just Tell Me, Ask Me: AI Systems that Intelligently Frame Explanations as Questions Improve Human Logical Discernment Accuracy over Causal AI explanations}. In \bibinfo{booktitle}{\emph{Proceedings of the 2023 CHI Conference on Human Factors in Computing Systems}}. \bibinfo{pages}{1--13}.
\newblock


\bibitem[de~Jong et~al\mbox{.}(2025)]%
        {de2025cognitive}
\bibfield{author}{\bibinfo{person}{Sander de Jong}, \bibinfo{person}{Ville Paananen}, \bibinfo{person}{Benjamin Tag}, {and} \bibinfo{person}{Niels van Berkel}.} \bibinfo{year}{2025}\natexlab{}.
\newblock \showarticletitle{Cognitive Forcing for Better Decision-Making: Reducing Overreliance on AI Systems Through Partial Explanations}.
\newblock \bibinfo{journal}{\emph{Proceedings of the ACM on Human-Computer Interaction-CSCW (2025)}} (\bibinfo{year}{2025}), \bibinfo{pages}{1--30}.
\newblock


\bibitem[Desender et~al\mbox{.}(2018)]%
        {desender2018subjective}
\bibfield{author}{\bibinfo{person}{Kobe Desender}, \bibinfo{person}{Annika Boldt}, {and} \bibinfo{person}{Nick Yeung}.} \bibinfo{year}{2018}\natexlab{}.
\newblock \showarticletitle{Subjective confidence predicts information seeking in decision making}.
\newblock \bibinfo{journal}{\emph{Psychological science}} \bibinfo{volume}{29}, \bibinfo{number}{5} (\bibinfo{year}{2018}), \bibinfo{pages}{761--778}.
\newblock


\bibitem[Ecker et~al\mbox{.}(2022)]%
        {ecker2022psychological}
\bibfield{author}{\bibinfo{person}{Ullrich~KH Ecker}, \bibinfo{person}{Stephan Lewandowsky}, \bibinfo{person}{John Cook}, \bibinfo{person}{Philipp Schmid}, \bibinfo{person}{Lisa~K Fazio}, \bibinfo{person}{Nadia Brashier}, \bibinfo{person}{Panayiota Kendeou}, \bibinfo{person}{Emily~K Vraga}, {and} \bibinfo{person}{Michelle~A Amazeen}.} \bibinfo{year}{2022}\natexlab{}.
\newblock \showarticletitle{The psychological drivers of misinformation belief and its resistance to correction}.
\newblock \bibinfo{journal}{\emph{Nature Reviews Psychology}} \bibinfo{volume}{1}, \bibinfo{number}{1} (\bibinfo{year}{2022}), \bibinfo{pages}{13--29}.
\newblock


\bibitem[Ehsan et~al\mbox{.}(2021a)]%
        {ehsan2021expanding}
\bibfield{author}{\bibinfo{person}{Upol Ehsan}, \bibinfo{person}{Q~Vera Liao}, \bibinfo{person}{Michael Muller}, \bibinfo{person}{Mark~O Riedl}, {and} \bibinfo{person}{Justin~D Weisz}.} \bibinfo{year}{2021}\natexlab{a}.
\newblock \showarticletitle{Expanding explainability: Towards social transparency in ai systems}. In \bibinfo{booktitle}{\emph{Proceedings of the 2021 CHI conference on human factors in computing systems}}. \bibinfo{pages}{1--19}.
\newblock


\bibitem[Ehsan et~al\mbox{.}(2021b)]%
        {ehsan2021explainable}
\bibfield{author}{\bibinfo{person}{Upol Ehsan}, \bibinfo{person}{Samir Passi}, \bibinfo{person}{Q~Vera Liao}, \bibinfo{person}{Larry Chan}, \bibinfo{person}{I Lee}, \bibinfo{person}{Michael Muller}, \bibinfo{person}{Mark~O Riedl}, {et~al\mbox{.}}} \bibinfo{year}{2021}\natexlab{b}.
\newblock \showarticletitle{The who in explainable ai: How ai background shapes perceptions of ai explanations}.
\newblock \bibinfo{journal}{\emph{arXiv preprint arXiv:2107.13509}} (\bibinfo{year}{2021}).
\newblock


\bibitem[Ehsan et~al\mbox{.}(2024)]%
        {ehsan2024xai}
\bibfield{author}{\bibinfo{person}{Upol Ehsan}, \bibinfo{person}{Samir Passi}, \bibinfo{person}{Q~Vera Liao}, \bibinfo{person}{Larry Chan}, \bibinfo{person}{I-Hsiang Lee}, \bibinfo{person}{Michael Muller}, {and} \bibinfo{person}{Mark~O Riedl}.} \bibinfo{year}{2024}\natexlab{}.
\newblock \showarticletitle{The Who in XAI: How AI Background Shapes Perceptions of AI Explanations}. In \bibinfo{booktitle}{\emph{Proceedings of the CHI Conference on Human Factors in Computing Systems}}. \bibinfo{pages}{1--32}.
\newblock


\bibitem[Ehsan et~al\mbox{.}(2019)]%
        {ehsan2019automated}
\bibfield{author}{\bibinfo{person}{Upol Ehsan}, \bibinfo{person}{Pradyumna Tambwekar}, \bibinfo{person}{Larry Chan}, \bibinfo{person}{Brent Harrison}, {and} \bibinfo{person}{Mark~O Riedl}.} \bibinfo{year}{2019}\natexlab{}.
\newblock \showarticletitle{Automated rationale generation: a technique for explainable AI and its effects on human perceptions}. In \bibinfo{booktitle}{\emph{Proceedings of the 24th international conference on intelligent user interfaces}}. \bibinfo{pages}{263--274}.
\newblock


\bibitem[Epstein et~al\mbox{.}(2022)]%
        {epstein2022explanations}
\bibfield{author}{\bibinfo{person}{Ziv Epstein}, \bibinfo{person}{Nicolo Foppiani}, \bibinfo{person}{Sophie Hilgard}, \bibinfo{person}{Sanjana Sharma}, \bibinfo{person}{Elena Glassman}, {and} \bibinfo{person}{David Rand}.} \bibinfo{year}{2022}\natexlab{}.
\newblock \showarticletitle{Do explanations increase the effectiveness of AI-crowd generated fake news warnings?}. In \bibinfo{booktitle}{\emph{Proceedings of the International AAAI Conference on Web and Social Media}}, Vol.~\bibinfo{volume}{16}. \bibinfo{pages}{183--193}.
\newblock


\bibitem[Gajos and Mamykina(2022)]%
        {gajos2022people}
\bibfield{author}{\bibinfo{person}{Krzysztof~Z Gajos} {and} \bibinfo{person}{Lena Mamykina}.} \bibinfo{year}{2022}\natexlab{}.
\newblock \showarticletitle{Do people engage cognitively with AI? Impact of AI assistance on incidental learning}. In \bibinfo{booktitle}{\emph{Proceedings of the 27th International Conference on Intelligent User Interfaces}}. \bibinfo{pages}{794--806}.
\newblock


\bibitem[Gangopadhyay et~al\mbox{.}(2024)]%
        {gangopadhyay2024investigating}
\bibfield{author}{\bibinfo{person}{Susmita Gangopadhyay}, \bibinfo{person}{Sebastian Schellhammer}, \bibinfo{person}{Salim Hafid}, \bibinfo{person}{Danilo Dessi}, \bibinfo{person}{Christian Ko{\ss}}, \bibinfo{person}{Konstantin Todorov}, \bibinfo{person}{Stefan Dietze}, {and} \bibinfo{person}{Hajira Jabeen}.} \bibinfo{year}{2024}\natexlab{}.
\newblock \showarticletitle{Investigating Characteristics, Biases and Evolution of Fact-Checked Claims on the Web}. In \bibinfo{booktitle}{\emph{Proceedings of the 35th ACM Conference on Hypertext and Social Media}}. \bibinfo{pages}{246--258}.
\newblock


\bibitem[Ghosh and Mitra(2023)]%
        {ghosh2023catching}
\bibfield{author}{\bibinfo{person}{Shreya Ghosh} {and} \bibinfo{person}{Prasenjit Mitra}.} \bibinfo{year}{2023}\natexlab{}.
\newblock \showarticletitle{Catching Lies in the Act: A Framework for Early Misinformation Detection on Social Media}. In \bibinfo{booktitle}{\emph{Proceedings of the 34th ACM Conference on Hypertext and Social Media}}. \bibinfo{pages}{1--12}.
\newblock


\bibitem[G{\"o}bel(2015)]%
        {gobel2015up}
\bibfield{author}{\bibinfo{person}{Silke~M G{\"o}bel}.} \bibinfo{year}{2015}\natexlab{}.
\newblock \showarticletitle{Up or down? Reading direction influences vertical counting direction in the horizontal plane--a cross-cultural comparison}.
\newblock \bibinfo{journal}{\emph{Frontiers in psychology}}  \bibinfo{volume}{6} (\bibinfo{year}{2015}), \bibinfo{pages}{228}.
\newblock


\bibitem[Gong et~al\mbox{.}(2024)]%
        {gong2024integrating}
\bibfield{author}{\bibinfo{person}{Yeaeun Gong}, \bibinfo{person}{Lanyu Shang}, {and} \bibinfo{person}{Dong Wang}.} \bibinfo{year}{2024}\natexlab{}.
\newblock \showarticletitle{Integrating Social Explanations into Explainable Artificial Intelligence (XAI) for Combating Misinformation: Vision and Challenges}.
\newblock \bibinfo{journal}{\emph{IEEE Transactions on Computational Social Systems (TCSS)}} (\bibinfo{year}{2024}).
\newblock


\bibitem[Gravanis et~al\mbox{.}(2019)]%
        {gravanis2019behind}
\bibfield{author}{\bibinfo{person}{Georgios Gravanis}, \bibinfo{person}{Athena Vakali}, \bibinfo{person}{Konstantinos Diamantaras}, {and} \bibinfo{person}{Panagiotis Karadais}.} \bibinfo{year}{2019}\natexlab{}.
\newblock \showarticletitle{Behind the cues: A benchmarking study for fake news detection}.
\newblock \bibinfo{journal}{\emph{Expert Systems with Applications}}  \bibinfo{volume}{128} (\bibinfo{year}{2019}), \bibinfo{pages}{201--213}.
\newblock


\bibitem[Guo et~al\mbox{.}(2018)]%
        {guo2018rumor}
\bibfield{author}{\bibinfo{person}{Han Guo}, \bibinfo{person}{Juan Cao}, \bibinfo{person}{Yazi Zhang}, \bibinfo{person}{Junbo Guo}, {and} \bibinfo{person}{Jintao Li}.} \bibinfo{year}{2018}\natexlab{}.
\newblock \showarticletitle{Rumor detection with hierarchical social attention network}. In \bibinfo{booktitle}{\emph{Proceedings of the 27th ACM international conference on information and knowledge management}}. \bibinfo{pages}{943--951}.
\newblock


\bibitem[Guo et~al\mbox{.}(2023)]%
        {guo2023two}
\bibfield{author}{\bibinfo{person}{Ying Guo}, \bibinfo{person}{Hong Ge}, {and} \bibinfo{person}{Jinhong Li}.} \bibinfo{year}{2023}\natexlab{}.
\newblock \showarticletitle{A two-branch multimodal fake news detection model based on multimodal bilinear pooling and attention mechanism}.
\newblock \bibinfo{journal}{\emph{Frontiers in Computer Science}}  \bibinfo{volume}{5} (\bibinfo{year}{2023}), \bibinfo{pages}{1159063}.
\newblock


\bibitem[Hart et~al\mbox{.}(2009)]%
        {hart2009feeling}
\bibfield{author}{\bibinfo{person}{William Hart}, \bibinfo{person}{Dolores Albarrac{\'\i}n}, \bibinfo{person}{Alice~H Eagly}, \bibinfo{person}{Inge Brechan}, \bibinfo{person}{Matthew~J Lindberg}, {and} \bibinfo{person}{Lisa Merrill}.} \bibinfo{year}{2009}\natexlab{}.
\newblock \showarticletitle{Feeling validated versus being correct: a meta-analysis of selective exposure to information.}
\newblock \bibinfo{journal}{\emph{Psychological bulletin}} \bibinfo{volume}{135}, \bibinfo{number}{4} (\bibinfo{year}{2009}), \bibinfo{pages}{555}.
\newblock


\bibitem[Hofmann et~al\mbox{.}(2024)]%
        {hofmann2024ai}
\bibfield{author}{\bibinfo{person}{Valentin Hofmann}, \bibinfo{person}{Pratyusha~Ria Kalluri}, \bibinfo{person}{Dan Jurafsky}, {and} \bibinfo{person}{Sharese King}.} \bibinfo{year}{2024}\natexlab{}.
\newblock \showarticletitle{AI generates covertly racist decisions about people based on their dialect}.
\newblock \bibinfo{journal}{\emph{Nature}} \bibinfo{volume}{633}, \bibinfo{number}{8028} (\bibinfo{year}{2024}), \bibinfo{pages}{147--154}.
\newblock


\bibitem[Horne(2024)]%
        {horne2024does}
\bibfield{author}{\bibinfo{person}{Benjamin~D Horne}.} \bibinfo{year}{2024}\natexlab{}.
\newblock \showarticletitle{Does the Source of a Warning Matter? Examining the Effectiveness of Veracity Warning Labels Across Warners}.
\newblock \bibinfo{journal}{\emph{arXiv preprint arXiv:2407.21592}} (\bibinfo{year}{2024}).
\newblock


\bibitem[Im et~al\mbox{.}(2020)]%
        {im2020synthesized}
\bibfield{author}{\bibinfo{person}{Jane Im}, \bibinfo{person}{Sonali Tandon}, \bibinfo{person}{Eshwar Chandrasekharan}, \bibinfo{person}{Taylor Denby}, {and} \bibinfo{person}{Eric Gilbert}.} \bibinfo{year}{2020}\natexlab{}.
\newblock \showarticletitle{Synthesized social signals: Computationally-derived social signals from account histories}. In \bibinfo{booktitle}{\emph{Proceedings of the 2020 CHI Conference on Human Factors in Computing Systems}}. \bibinfo{pages}{1--12}.
\newblock


\bibitem[Jacobs et~al\mbox{.}(2021)]%
        {jacobs2021designing}
\bibfield{author}{\bibinfo{person}{Maia Jacobs}, \bibinfo{person}{Jeffrey He}, \bibinfo{person}{Melanie F.~Pradier}, \bibinfo{person}{Barbara Lam}, \bibinfo{person}{Andrew~C Ahn}, \bibinfo{person}{Thomas~H McCoy}, \bibinfo{person}{Roy~H Perlis}, \bibinfo{person}{Finale Doshi-Velez}, {and} \bibinfo{person}{Krzysztof~Z Gajos}.} \bibinfo{year}{2021}\natexlab{}.
\newblock \showarticletitle{Designing AI for trust and collaboration in time-constrained medical decisions: a sociotechnical lens}. In \bibinfo{booktitle}{\emph{Proceedings of the 2021 chi conference on human factors in computing systems}}. \bibinfo{pages}{1--14}.
\newblock


\bibitem[Jambor and Bornh{\"a}user(2024)]%
        {jambor2024ten}
\bibfield{author}{\bibinfo{person}{Helena~Klara Jambor} {and} \bibinfo{person}{Martin Bornh{\"a}user}.} \bibinfo{year}{2024}\natexlab{}.
\newblock \showarticletitle{Ten simple rules for designing graphical abstracts}.
\newblock \bibinfo{journal}{\emph{PLOS Computational Biology}} \bibinfo{volume}{20}, \bibinfo{number}{2} (\bibinfo{year}{2024}), \bibinfo{pages}{e1011789}.
\newblock


\bibitem[Jeronimo et~al\mbox{.}(2019)]%
        {jeronimo2019fake}
\bibfield{author}{\bibinfo{person}{Caio Libanio~Melo Jeronimo}, \bibinfo{person}{Leandro~Balby Marinho}, \bibinfo{person}{Claudio~EC Campelo}, \bibinfo{person}{Adriano Veloso}, {and} \bibinfo{person}{Allan~Sales da Costa~Melo}.} \bibinfo{year}{2019}\natexlab{}.
\newblock \showarticletitle{Fake news classification based on subjective language}. In \bibinfo{booktitle}{\emph{Proceedings of the 21st International Conference on Information Integration and Web-based Applications \& Services}}. \bibinfo{pages}{15--24}.
\newblock


\bibitem[Joy et~al\mbox{.}(2021)]%
        {joy2021you}
\bibfield{author}{\bibinfo{person}{Abishai Joy}, \bibinfo{person}{Anu Shrestha}, {and} \bibinfo{person}{Francesca Spezzano}.} \bibinfo{year}{2021}\natexlab{}.
\newblock \showarticletitle{Are you influenced? modeling the diffusion of fake news in social media}. In \bibinfo{booktitle}{\emph{Proceedings of the 2021 IEEE/ACM International Conference on Advances in Social Networks Analysis and Mining}}. \bibinfo{pages}{184--188}.
\newblock


\bibitem[Klayman(1995)]%
        {klayman1995varieties}
\bibfield{author}{\bibinfo{person}{Joshua Klayman}.} \bibinfo{year}{1995}\natexlab{}.
\newblock \showarticletitle{Varieties of confirmation bias}.
\newblock \bibinfo{journal}{\emph{Psychology of learning and motivation}}  \bibinfo{volume}{32} (\bibinfo{year}{1995}), \bibinfo{pages}{385--418}.
\newblock


\bibitem[Kou et~al\mbox{.}(2022a)]%
        {kou2022hc}
\bibfield{author}{\bibinfo{person}{Ziyi Kou}, \bibinfo{person}{Lanyu Shang}, \bibinfo{person}{Yang Zhang}, {and} \bibinfo{person}{Dong Wang}.} \bibinfo{year}{2022}\natexlab{a}.
\newblock \showarticletitle{Hc-covid: A hierarchical crowdsource knowledge graph approach to explainable covid-19 misinformation detection}.
\newblock \bibinfo{journal}{\emph{Proceedings of the ACM on Human-Computer Interaction}} \bibinfo{volume}{6}, \bibinfo{number}{GROUP} (\bibinfo{year}{2022}), \bibinfo{pages}{1--25}.
\newblock


\bibitem[Kou et~al\mbox{.}(2022b)]%
        {kou2022crowd}
\bibfield{author}{\bibinfo{person}{Ziyi Kou}, \bibinfo{person}{Lanyu Shang}, \bibinfo{person}{Yang Zhang}, \bibinfo{person}{Zhenrui Yue}, \bibinfo{person}{Huimin Zeng}, {and} \bibinfo{person}{Dong Wang}.} \bibinfo{year}{2022}\natexlab{b}.
\newblock \showarticletitle{Crowd, expert \& ai: A human-ai interactive approach towards natural language explanation based covid-19 misinformation detection}. In \bibinfo{booktitle}{\emph{Proc. Int. Joint Conf. Artif. Intell.(IJCAI)}}. \bibinfo{pages}{5087–5093}.
\newblock


\bibitem[Kumar and Taylor(2024)]%
        {kumar2024feature}
\bibfield{author}{\bibinfo{person}{Ajay Kumar} {and} \bibinfo{person}{James~W Taylor}.} \bibinfo{year}{2024}\natexlab{}.
\newblock \showarticletitle{Feature importance in the age of explainable AI: Case study of detecting fake news \& misinformation via a multi-modal framework}.
\newblock \bibinfo{journal}{\emph{European Journal of Operational Research}} \bibinfo{volume}{317}, \bibinfo{number}{2} (\bibinfo{year}{2024}), \bibinfo{pages}{401--413}.
\newblock


\bibitem[Lai et~al\mbox{.}(2023)]%
        {lai2023selective}
\bibfield{author}{\bibinfo{person}{Vivian Lai}, \bibinfo{person}{Yiming Zhang}, \bibinfo{person}{Chacha Chen}, \bibinfo{person}{Q~Vera Liao}, {and} \bibinfo{person}{Chenhao Tan}.} \bibinfo{year}{2023}\natexlab{}.
\newblock \showarticletitle{Selective Explanations: Leveraging Human Input to Align Explainable AI}.
\newblock \bibinfo{journal}{\emph{Proceedings of the ACM on Human-Computer Interaction}} (\bibinfo{year}{2023}).
\newblock


\bibitem[Lee et~al\mbox{.}(2022)]%
        {lee2022community}
\bibfield{author}{\bibinfo{person}{Jin~Ha Lee}, \bibinfo{person}{Nicole Santero}, \bibinfo{person}{Arpita Bhattacharya}, \bibinfo{person}{Emma May}, {and} \bibinfo{person}{Emma~S Spiro}.} \bibinfo{year}{2022}\natexlab{}.
\newblock \showarticletitle{Community-based strategies for combating misinformation: Learning from a popular culture fandom}.
\newblock \bibinfo{journal}{\emph{Harvard Kennedy School Misinformation Review}} (\bibinfo{year}{2022}).
\newblock


\bibitem[Lee and Chew(2023)]%
        {lee2023understanding}
\bibfield{author}{\bibinfo{person}{Min~Hun Lee} {and} \bibinfo{person}{Chong~Jun Chew}.} \bibinfo{year}{2023}\natexlab{}.
\newblock \showarticletitle{Understanding the effect of counterfactual explanations on trust and reliance on ai for human-ai collaborative clinical decision making}.
\newblock \bibinfo{journal}{\emph{Proceedings of the ACM on Human-Computer Interaction}} \bibinfo{volume}{7}, \bibinfo{number}{CSCW2} (\bibinfo{year}{2023}), \bibinfo{pages}{1--22}.
\newblock


\bibitem[Li et~al\mbox{.}(2023)]%
        {li2023assessing}
\bibfield{author}{\bibinfo{person}{Junhao Li}, \bibinfo{person}{Miikka Kuutila}, \bibinfo{person}{Eetu Huusko}, \bibinfo{person}{Nimantha Kariyakarawana}, \bibinfo{person}{Marko Savic}, \bibinfo{person}{Nazanin~Nakhaie Ahooie}, \bibinfo{person}{Simo Hosio}, {and} \bibinfo{person}{Mika M{\"a}ntyl{\"a}}.} \bibinfo{year}{2023}\natexlab{}.
\newblock \showarticletitle{Assessing credibility factors of short-form social media posts: A crowdsourced online experiment}. In \bibinfo{booktitle}{\emph{Proceedings of the 15th biannual conference of the Italian SIGCHI chapter}}. \bibinfo{pages}{1--14}.
\newblock


\bibitem[Liao and Varshney(2021)]%
        {liao2021human}
\bibfield{author}{\bibinfo{person}{Q~Vera Liao} {and} \bibinfo{person}{Kush~R Varshney}.} \bibinfo{year}{2021}\natexlab{}.
\newblock \showarticletitle{Human-centered explainable ai (xai): From algorithms to user experiences}.
\newblock \bibinfo{journal}{\emph{arXiv preprint arXiv:2110.10790}} (\bibinfo{year}{2021}).
\newblock


\bibitem[Liao and Vaughan(2023)]%
        {liao2023ai}
\bibfield{author}{\bibinfo{person}{Q~Vera Liao} {and} \bibinfo{person}{Jennifer~Wortman Vaughan}.} \bibinfo{year}{2023}\natexlab{}.
\newblock \showarticletitle{AI Transparency in the Age of LLMs: A Human-Centered Research Roadmap}.
\newblock \bibinfo{journal}{\emph{arXiv preprint arXiv:2306.01941}} (\bibinfo{year}{2023}).
\newblock


\bibitem[Lin et~al\mbox{.}(2022)]%
        {lin2022teaching}
\bibfield{author}{\bibinfo{person}{Stephanie Lin}, \bibinfo{person}{Jacob Hilton}, {and} \bibinfo{person}{Owain Evans}.} \bibinfo{year}{2022}\natexlab{}.
\newblock \showarticletitle{Teaching models to express their uncertainty in words}.
\newblock \bibinfo{journal}{\emph{arXiv preprint arXiv:2205.14334}} (\bibinfo{year}{2022}).
\newblock


\bibitem[Linder et~al\mbox{.}(2021)]%
        {linder2021level}
\bibfield{author}{\bibinfo{person}{Rhema Linder}, \bibinfo{person}{Sina Mohseni}, \bibinfo{person}{Fan Yang}, \bibinfo{person}{Shiva~K Pentyala}, \bibinfo{person}{Eric~D Ragan}, {and} \bibinfo{person}{Xia~Ben Hu}.} \bibinfo{year}{2021}\natexlab{}.
\newblock \showarticletitle{How level of explanation detail affects human performance in interpretable intelligent systems: A study on explainable fact checking}.
\newblock \bibinfo{journal}{\emph{Applied AI Letters}} \bibinfo{volume}{2}, \bibinfo{number}{4} (\bibinfo{year}{2021}), \bibinfo{pages}{e49}.
\newblock


\bibitem[Liu and Wu(2018)]%
        {liu2018early}
\bibfield{author}{\bibinfo{person}{Yang Liu} {and} \bibinfo{person}{Yi-Fang Wu}.} \bibinfo{year}{2018}\natexlab{}.
\newblock \showarticletitle{Early detection of fake news on social media through propagation path classification with recurrent and convolutional networks}. In \bibinfo{booktitle}{\emph{Proceedings of the AAAI conference on artificial intelligence}}, Vol.~\bibinfo{volume}{32}.
\newblock


\bibitem[Lodge et~al\mbox{.}(2018)]%
        {lodge2018understanding}
\bibfield{author}{\bibinfo{person}{Jason~M Lodge}, \bibinfo{person}{Gregor Kennedy}, \bibinfo{person}{Lori Lockyer}, \bibinfo{person}{Amael Arguel}, {and} \bibinfo{person}{Mariya Pachman}.} \bibinfo{year}{2018}\natexlab{}.
\newblock \showarticletitle{Understanding difficulties and resulting confusion in learning: An integrative review}. In \bibinfo{booktitle}{\emph{Frontiers in Education}}, Vol.~\bibinfo{volume}{3}. Frontiers Media SA, \bibinfo{pages}{49}.
\newblock


\bibitem[Long et~al\mbox{.}(2017)]%
        {long2017fake}
\bibfield{author}{\bibinfo{person}{Yunfei Long}, \bibinfo{person}{Qin Lu}, \bibinfo{person}{Rong Xiang}, \bibinfo{person}{Minglei Li}, {and} \bibinfo{person}{Chu-Ren Huang}.} \bibinfo{year}{2017}\natexlab{}.
\newblock \showarticletitle{Fake news detection through multi-perspective speaker profiles}. In \bibinfo{booktitle}{\emph{Proceedings of the eighth international joint conference on natural language processing (volume 2: Short papers)}}. \bibinfo{pages}{252--256}.
\newblock


\bibitem[Michael and Breaux(2021)]%
        {michael2021relationship}
\bibfield{author}{\bibinfo{person}{Robert~B Michael} {and} \bibinfo{person}{Brooke~O Breaux}.} \bibinfo{year}{2021}\natexlab{}.
\newblock \showarticletitle{The relationship between political affiliation and beliefs about sources of “fake news”}.
\newblock \bibinfo{journal}{\emph{Cognitive research: principles and implications}}  \bibinfo{volume}{6} (\bibinfo{year}{2021}), \bibinfo{pages}{1--15}.
\newblock


\bibitem[Mohseni et~al\mbox{.}(2021)]%
        {mohseni2021machine}
\bibfield{author}{\bibinfo{person}{Sina Mohseni}, \bibinfo{person}{Fan Yang}, \bibinfo{person}{Shiva Pentyala}, \bibinfo{person}{Mengnan Du}, \bibinfo{person}{Yi Liu}, \bibinfo{person}{Nic Lupfer}, \bibinfo{person}{Xia Hu}, \bibinfo{person}{Shuiwang Ji}, {and} \bibinfo{person}{Eric Ragan}.} \bibinfo{year}{2021}\natexlab{}.
\newblock \showarticletitle{Machine learning explanations to prevent overtrust in fake news detection}. In \bibinfo{booktitle}{\emph{Proceedings of the international AAAI conference on web and social media}}, Vol.~\bibinfo{volume}{15}. \bibinfo{pages}{421--431}.
\newblock


\bibitem[Mondak(1993)]%
        {mondak1993public}
\bibfield{author}{\bibinfo{person}{Jeffery~J Mondak}.} \bibinfo{year}{1993}\natexlab{}.
\newblock \showarticletitle{Public opinion and heuristic processing of source cues}.
\newblock \bibinfo{journal}{\emph{Political behavior}}  \bibinfo{volume}{15} (\bibinfo{year}{1993}), \bibinfo{pages}{167--192}.
\newblock


\bibitem[Morrison et~al\mbox{.}(2024)]%
        {morrison2024impact}
\bibfield{author}{\bibinfo{person}{Katelyn Morrison}, \bibinfo{person}{Philipp Spitzer}, \bibinfo{person}{Violet Turri}, \bibinfo{person}{Michelle Feng}, \bibinfo{person}{Niklas K{\"u}hl}, {and} \bibinfo{person}{Adam Perer}.} \bibinfo{year}{2024}\natexlab{}.
\newblock \showarticletitle{The Impact of Imperfect XAI on Human-AI Decision-Making}.
\newblock \bibinfo{journal}{\emph{Proceedings of the ACM on Human-Computer Interaction}} \bibinfo{volume}{8}, \bibinfo{number}{CSCW1} (\bibinfo{year}{2024}), \bibinfo{pages}{1–39}.
\newblock


\bibitem[Mosleh and Rand(2022)]%
        {mosleh2022measuring}
\bibfield{author}{\bibinfo{person}{Mohsen Mosleh} {and} \bibinfo{person}{David~G Rand}.} \bibinfo{year}{2022}\natexlab{}.
\newblock \showarticletitle{Measuring exposure to misinformation from political elites on Twitter}.
\newblock \bibinfo{journal}{\emph{Nature Communications}} \bibinfo{volume}{13}, \bibinfo{number}{1} (\bibinfo{year}{2022}), \bibinfo{pages}{7144}.
\newblock


\bibitem[Muhammed~T and Mathew(2022)]%
        {muhammed2022disaster}
\bibfield{author}{\bibinfo{person}{Sadiq Muhammed~T} {and} \bibinfo{person}{Saji~K Mathew}.} \bibinfo{year}{2022}\natexlab{}.
\newblock \showarticletitle{The disaster of misinformation: a review of research in social media}.
\newblock \bibinfo{journal}{\emph{International journal of data science and analytics}} \bibinfo{volume}{13}, \bibinfo{number}{4} (\bibinfo{year}{2022}), \bibinfo{pages}{271--285}.
\newblock


\bibitem[Pafla et~al\mbox{.}(2024)]%
        {pafla2024unraveling}
\bibfield{author}{\bibinfo{person}{Marvin Pafla}, \bibinfo{person}{Kate Larson}, {and} \bibinfo{person}{Mark Hancock}.} \bibinfo{year}{2024}\natexlab{}.
\newblock \showarticletitle{Unraveling the Dilemma of AI Errors: Exploring the Effectiveness of Human and Machine Explanations for Large Language Models}. In \bibinfo{booktitle}{\emph{Proceedings of the CHI Conference on Human Factors in Computing Systems}}. \bibinfo{pages}{1--20}.
\newblock


\bibitem[Panigutti et~al\mbox{.}(2022)]%
        {panigutti2022understanding}
\bibfield{author}{\bibinfo{person}{Cecilia Panigutti}, \bibinfo{person}{Andrea Beretta}, \bibinfo{person}{Fosca Giannotti}, {and} \bibinfo{person}{Dino Pedreschi}.} \bibinfo{year}{2022}\natexlab{}.
\newblock \showarticletitle{Understanding the impact of explanations on advice-taking: a user study for AI-based clinical Decision Support Systems}. In \bibinfo{booktitle}{\emph{Proceedings of the 2022 CHI Conference on Human Factors in Computing Systems}}. \bibinfo{pages}{1--9}.
\newblock


\bibitem[Pennycook et~al\mbox{.}(2021)]%
        {pennycook2021shifting}
\bibfield{author}{\bibinfo{person}{Gordon Pennycook}, \bibinfo{person}{Ziv Epstein}, \bibinfo{person}{Mohsen Mosleh}, \bibinfo{person}{Antonio~A Arechar}, \bibinfo{person}{Dean Eckles}, {and} \bibinfo{person}{David~G Rand}.} \bibinfo{year}{2021}\natexlab{}.
\newblock \showarticletitle{Shifting attention to accuracy can reduce misinformation online}.
\newblock \bibinfo{journal}{\emph{Nature}} \bibinfo{volume}{592}, \bibinfo{number}{7855} (\bibinfo{year}{2021}), \bibinfo{pages}{590--595}.
\newblock


\bibitem[Petroni et~al\mbox{.}(2019)]%
        {petroni2019language}
\bibfield{author}{\bibinfo{person}{Fabio Petroni}, \bibinfo{person}{Tim Rockt{\"a}schel}, \bibinfo{person}{Patrick Lewis}, \bibinfo{person}{Anton Bakhtin}, \bibinfo{person}{Yuxiang Wu}, \bibinfo{person}{Alexander~H Miller}, {and} \bibinfo{person}{Sebastian Riedel}.} \bibinfo{year}{2019}\natexlab{}.
\newblock \showarticletitle{Language models as knowledge bases?}
\newblock \bibinfo{journal}{\emph{arXiv preprint arXiv:1909.01066}} (\bibinfo{year}{2019}).
\newblock


\bibitem[Prakash and Kumar(2023)]%
        {prakash2023fake}
\bibfield{author}{\bibinfo{person}{Om Prakash} {and} \bibinfo{person}{Rajeev Kumar}.} \bibinfo{year}{2023}\natexlab{}.
\newblock \showarticletitle{Fake news detection in social networks using attention mechanism}. In \bibinfo{booktitle}{\emph{Proceedings of the International Conference on Cognitive and Intelligent Computing: ICCIC 2021, Volume 2}}. Springer, \bibinfo{pages}{453--462}.
\newblock


\bibitem[Purificato et~al\mbox{.}(2022)]%
        {purificato2022tell}
\bibfield{author}{\bibinfo{person}{Erasmo Purificato}, \bibinfo{person}{Saijal Shahania}, {and} \bibinfo{person}{Ernesto~William De~Luca}.} \bibinfo{year}{2022}\natexlab{}.
\newblock \showarticletitle{Tell Me Why It's Fake: Developing an Explainable User Interface for a Fake News Detection System.}. In \bibinfo{booktitle}{\emph{XAI. it@ AI* IA}}. \bibinfo{pages}{51--63}.
\newblock


\bibitem[Radford et~al\mbox{.}(2019)]%
        {radford2019language}
\bibfield{author}{\bibinfo{person}{Alec Radford}, \bibinfo{person}{Jeffrey Wu}, \bibinfo{person}{Rewon Child}, \bibinfo{person}{David Luan}, \bibinfo{person}{Dario Amodei}, \bibinfo{person}{Ilya Sutskever}, {et~al\mbox{.}}} \bibinfo{year}{2019}\natexlab{}.
\newblock \showarticletitle{Language models are unsupervised multitask learners}.
\newblock \bibinfo{journal}{\emph{OpenAI blog}} \bibinfo{volume}{1}, \bibinfo{number}{8} (\bibinfo{year}{2019}), \bibinfo{pages}{9}.
\newblock


\bibitem[Rangapur et~al\mbox{.}(2023)]%
        {rangapur2023investigating}
\bibfield{author}{\bibinfo{person}{Aman Rangapur}, \bibinfo{person}{Haoran Wang}, {and} \bibinfo{person}{Kai Shu}.} \bibinfo{year}{2023}\natexlab{}.
\newblock \showarticletitle{Investigating online financial misinformation and its consequences: A computational perspective}.
\newblock \bibinfo{journal}{\emph{arXiv preprint arXiv:2309.12363}} (\bibinfo{year}{2023}).
\newblock


\bibitem[Rechkemmer and Yin(2022)]%
        {rechkemmer2022confidence}
\bibfield{author}{\bibinfo{person}{Amy Rechkemmer} {and} \bibinfo{person}{Ming Yin}.} \bibinfo{year}{2022}\natexlab{}.
\newblock \showarticletitle{When confidence meets accuracy: Exploring the effects of multiple performance indicators on trust in machine learning models}. In \bibinfo{booktitle}{\emph{Proceedings of the 2022 chi conference on human factors in computing systems}}. \bibinfo{pages}{1--14}.
\newblock


\bibitem[Reinhard and Sporer(2010)]%
        {reinhard2010content}
\bibfield{author}{\bibinfo{person}{Marc-Andr{\'e} Reinhard} {and} \bibinfo{person}{Siegfried~L Sporer}.} \bibinfo{year}{2010}\natexlab{}.
\newblock \showarticletitle{Content versus source cue information as a basis for credibility judgments}.
\newblock \bibinfo{journal}{\emph{Social Psychology}} (\bibinfo{year}{2010}).
\newblock


\bibitem[Roozenbeek et~al\mbox{.}(2022)]%
        {roozenbeek2022susceptibility}
\bibfield{author}{\bibinfo{person}{Jon Roozenbeek}, \bibinfo{person}{Rakoen Maertens}, \bibinfo{person}{Stefan~M Herzog}, \bibinfo{person}{Michael Geers}, \bibinfo{person}{Ralf Kurvers}, \bibinfo{person}{Mubashir Sultan}, {and} \bibinfo{person}{Sander van~der Linden}.} \bibinfo{year}{2022}\natexlab{}.
\newblock \showarticletitle{Susceptibility to misinformation is consistent across questionframings and response modes and better explained by myside bias and partisanshipthan analytical thinking}.
\newblock \bibinfo{journal}{\emph{Judgment and Decision Making}} \bibinfo{volume}{17}, \bibinfo{number}{3} (\bibinfo{year}{2022}), \bibinfo{pages}{547--573}.
\newblock


\bibitem[Russo et~al\mbox{.}(2008)]%
        {russo2008goal}
\bibfield{author}{\bibinfo{person}{J~Edward Russo}, \bibinfo{person}{Kurt~A Carlson}, \bibinfo{person}{Margaret~G Meloy}, {and} \bibinfo{person}{Kevyn Yong}.} \bibinfo{year}{2008}\natexlab{}.
\newblock \showarticletitle{The goal of consistency as a cause of information distortion.}
\newblock \bibinfo{journal}{\emph{Journal of Experimental Psychology: General}} \bibinfo{volume}{137}, \bibinfo{number}{3} (\bibinfo{year}{2008}), \bibinfo{pages}{456}.
\newblock


\bibitem[Schwind and Buder(2012)]%
        {schwind2012reducing}
\bibfield{author}{\bibinfo{person}{Christina Schwind} {and} \bibinfo{person}{J{\"u}rgen Buder}.} \bibinfo{year}{2012}\natexlab{}.
\newblock \showarticletitle{Reducing confirmation bias and evaluation bias: When are preference-inconsistent recommendations effective--and when not?}
\newblock \bibinfo{journal}{\emph{Computers in Human Behavior}} \bibinfo{volume}{28}, \bibinfo{number}{6} (\bibinfo{year}{2012}), \bibinfo{pages}{2280--2290}.
\newblock


\bibitem[Seo et~al\mbox{.}(2024)]%
        {seo2024reliability}
\bibfield{author}{\bibinfo{person}{Haeseung Seo}, \bibinfo{person}{Sian Lee}, \bibinfo{person}{Dongwon Lee}, {and} \bibinfo{person}{Aiping Xiong}.} \bibinfo{year}{2024}\natexlab{}.
\newblock \showarticletitle{Reliability Matters: Exploring the Effect of AI Explanations on Misinformation Detection with a Warning}. In \bibinfo{booktitle}{\emph{Proceedings of the International AAAI Conference on Web and Social Media}}, Vol.~\bibinfo{volume}{18}. \bibinfo{pages}{1395--1407}.
\newblock


\bibitem[Shang et~al\mbox{.}(2022)]%
        {shang2022duo}
\bibfield{author}{\bibinfo{person}{Lanyu Shang}, \bibinfo{person}{Ziyi Kou}, \bibinfo{person}{Yang Zhang}, {and} \bibinfo{person}{Dong Wang}.} \bibinfo{year}{2022}\natexlab{}.
\newblock \showarticletitle{A duo-generative approach to explainable multimodal covid-19 misinformation detection}. In \bibinfo{booktitle}{\emph{Proceedings of the ACM Web Conference 2022}}. \bibinfo{pages}{3623--3631}.
\newblock


\bibitem[Shao et~al\mbox{.}(2018)]%
        {shao2018spread}
\bibfield{author}{\bibinfo{person}{Chengcheng Shao}, \bibinfo{person}{Giovanni~Luca Ciampaglia}, \bibinfo{person}{Onur Varol}, \bibinfo{person}{Kai-Cheng Yang}, \bibinfo{person}{Alessandro Flammini}, {and} \bibinfo{person}{Filippo Menczer}.} \bibinfo{year}{2018}\natexlab{}.
\newblock \showarticletitle{The spread of low-credibility content by social bots}.
\newblock \bibinfo{journal}{\emph{Nature communications}} \bibinfo{volume}{9}, \bibinfo{number}{1} (\bibinfo{year}{2018}), \bibinfo{pages}{4787}.
\newblock


\bibitem[Shu et~al\mbox{.}(2017)]%
        {shu2017fake}
\bibfield{author}{\bibinfo{person}{Kai Shu}, \bibinfo{person}{Amy Sliva}, \bibinfo{person}{Suhang Wang}, \bibinfo{person}{Jiliang Tang}, {and} \bibinfo{person}{Huan Liu}.} \bibinfo{year}{2017}\natexlab{}.
\newblock \showarticletitle{Fake news detection on social media: A data mining perspective}.
\newblock \bibinfo{journal}{\emph{ACM SIGKDD explorations newsletter}} \bibinfo{volume}{19}, \bibinfo{number}{1} (\bibinfo{year}{2017}), \bibinfo{pages}{22--36}.
\newblock


\bibitem[Shu et~al\mbox{.}(2019a)]%
        {shu2019beyond}
\bibfield{author}{\bibinfo{person}{Kai Shu}, \bibinfo{person}{Suhang Wang}, {and} \bibinfo{person}{Huan Liu}.} \bibinfo{year}{2019}\natexlab{a}.
\newblock \showarticletitle{Beyond news contents: The role of social context for fake news detection}. In \bibinfo{booktitle}{\emph{Proceedings of the twelfth ACM international conference on web search and data mining}}. \bibinfo{pages}{312--320}.
\newblock


\bibitem[Shu et~al\mbox{.}(2019b)]%
        {shu2019role}
\bibfield{author}{\bibinfo{person}{Kai Shu}, \bibinfo{person}{Xinyi Zhou}, \bibinfo{person}{Suhang Wang}, \bibinfo{person}{Reza Zafarani}, {and} \bibinfo{person}{Huan Liu}.} \bibinfo{year}{2019}\natexlab{b}.
\newblock \showarticletitle{The role of user profiles for fake news detection}. In \bibinfo{booktitle}{\emph{Proceedings of the 2019 IEEE/ACM international conference on advances in social networks analysis and mining}}. \bibinfo{pages}{436--439}.
\newblock


\bibitem[Springer and Whittaker(2019)]%
        {springer2019progressive}
\bibfield{author}{\bibinfo{person}{Aaron Springer} {and} \bibinfo{person}{Steve Whittaker}.} \bibinfo{year}{2019}\natexlab{}.
\newblock \showarticletitle{Progressive disclosure: empirically motivated approaches to designing effective transparency}. In \bibinfo{booktitle}{\emph{Proceedings of the 24th international conference on intelligent user interfaces}}. \bibinfo{pages}{107--120}.
\newblock


\bibitem[Swire et~al\mbox{.}(2017)]%
        {swire2017processing}
\bibfield{author}{\bibinfo{person}{Briony Swire}, \bibinfo{person}{Adam~J Berinsky}, \bibinfo{person}{Stephan Lewandowsky}, {and} \bibinfo{person}{Ullrich~KH Ecker}.} \bibinfo{year}{2017}\natexlab{}.
\newblock \showarticletitle{Processing political misinformation: Comprehending the Trump phenomenon}.
\newblock \bibinfo{journal}{\emph{Royal Society open science}} \bibinfo{volume}{4}, \bibinfo{number}{3} (\bibinfo{year}{2017}), \bibinfo{pages}{160802}.
\newblock


\bibitem[Talluri et~al\mbox{.}(2018)]%
        {talluri2018confirmation}
\bibfield{author}{\bibinfo{person}{Bharath~Chandra Talluri}, \bibinfo{person}{Anne~E Urai}, \bibinfo{person}{Konstantinos Tsetsos}, \bibinfo{person}{Marius Usher}, {and} \bibinfo{person}{Tobias~H Donner}.} \bibinfo{year}{2018}\natexlab{}.
\newblock \showarticletitle{Confirmation bias through selective overweighting of choice-consistent evidence}.
\newblock \bibinfo{journal}{\emph{Current Biology}} \bibinfo{volume}{28}, \bibinfo{number}{19} (\bibinfo{year}{2018}), \bibinfo{pages}{3128--3135}.
\newblock


\bibitem[Turner et~al\mbox{.}(2010)]%
        {turner2010lay}
\bibfield{author}{\bibinfo{person}{Monique~Mitchell Turner}, \bibinfo{person}{Shuo Yao}, \bibinfo{person}{Ryanne Baker}, \bibinfo{person}{Jodi Goodman}, {and} \bibinfo{person}{Stephanie~A Materese}.} \bibinfo{year}{2010}\natexlab{}.
\newblock \showarticletitle{Do lay people prepare both sides of an argument? The effects of confidence, forewarning, and expected interaction on seeking out counter-attitudinal information}.
\newblock \bibinfo{journal}{\emph{Argumentation and Advocacy}} \bibinfo{volume}{46}, \bibinfo{number}{4} (\bibinfo{year}{2010}), \bibinfo{pages}{226--239}.
\newblock


\bibitem[Vasconcelos et~al\mbox{.}(2023)]%
        {vasconcelos2023explanations}
\bibfield{author}{\bibinfo{person}{Helena Vasconcelos}, \bibinfo{person}{Matthew J{\"o}rke}, \bibinfo{person}{Madeleine Grunde-McLaughlin}, \bibinfo{person}{Tobias Gerstenberg}, \bibinfo{person}{Michael~S Bernstein}, {and} \bibinfo{person}{Ranjay Krishna}.} \bibinfo{year}{2023}\natexlab{}.
\newblock \showarticletitle{Explanations can reduce overreliance on ai systems during decision-making}.
\newblock \bibinfo{journal}{\emph{Proceedings of the ACM on Human-Computer Interaction}} \bibinfo{volume}{7}, \bibinfo{number}{CSCW1} (\bibinfo{year}{2023}), \bibinfo{pages}{1--38}.
\newblock


\bibitem[Vieira et~al\mbox{.}(2020)]%
        {vieira2020analysis}
\bibfield{author}{\bibinfo{person}{Lucas~Lima Vieira}, \bibinfo{person}{Caio Libanio~Melo Jeronimo}, \bibinfo{person}{Claudio~EC Campelo}, {and} \bibinfo{person}{Leandro~Balby Marinho}.} \bibinfo{year}{2020}\natexlab{}.
\newblock \showarticletitle{Analysis of the subjectivity level in fake news fragments}. In \bibinfo{booktitle}{\emph{Proceedings of the Brazilian Symposium on Multimedia and the Web}}. \bibinfo{pages}{233--240}.
\newblock


\bibitem[Vosoughi et~al\mbox{.}(2018)]%
        {vosoughi2018spread}
\bibfield{author}{\bibinfo{person}{Soroush Vosoughi}, \bibinfo{person}{Deb Roy}, {and} \bibinfo{person}{Sinan Aral}.} \bibinfo{year}{2018}\natexlab{}.
\newblock \showarticletitle{The spread of true and false news online}.
\newblock \bibinfo{journal}{\emph{science}} \bibinfo{volume}{359}, \bibinfo{number}{6380} (\bibinfo{year}{2018}), \bibinfo{pages}{1146--1151}.
\newblock


\bibitem[Waltenberger et~al\mbox{.}(2023)]%
        {waltenberger2023reddit}
\bibfield{author}{\bibinfo{person}{Franz Waltenberger}, \bibinfo{person}{Simon H{\"o}ferlin}, {and} \bibinfo{person}{Michael Froehlich}.} \bibinfo{year}{2023}\natexlab{}.
\newblock \showarticletitle{Reddit Insights: Improving Online Discussion Culture by Contextualizing User Profiles}. In \bibinfo{booktitle}{\emph{Extended Abstracts of the 2023 CHI Conference on Human Factors in Computing Systems}}. \bibinfo{pages}{1--6}.
\newblock


\bibitem[Wang(2017)]%
        {wang2017liar}
\bibfield{author}{\bibinfo{person}{William~Yang Wang}.} \bibinfo{year}{2017}\natexlab{}.
\newblock \showarticletitle{" liar, liar pants on fire": A new benchmark dataset for fake news detection}.
\newblock \bibinfo{journal}{\emph{arXiv preprint arXiv:1705.00648}} (\bibinfo{year}{2017}).
\newblock


\bibitem[Wang et~al\mbox{.}(2024)]%
        {wang2024human}
\bibfield{author}{\bibinfo{person}{Xinru Wang}, \bibinfo{person}{Hannah Kim}, \bibinfo{person}{Sajjadur Rahman}, \bibinfo{person}{Kushan Mitra}, {and} \bibinfo{person}{Zhengjie Miao}.} \bibinfo{year}{2024}\natexlab{}.
\newblock \showarticletitle{Human-LLM collaborative annotation through effective verification of LLM labels}. In \bibinfo{booktitle}{\emph{Proceedings of the CHI Conference on Human Factors in Computing Systems}}. \bibinfo{pages}{1--21}.
\newblock


\bibitem[Wang and Yin(2021)]%
        {wang2021explanations}
\bibfield{author}{\bibinfo{person}{Xinru Wang} {and} \bibinfo{person}{Ming Yin}.} \bibinfo{year}{2021}\natexlab{}.
\newblock \showarticletitle{Are explanations helpful? a comparative study of the effects of explanations in ai-assisted decision-making}. In \bibinfo{booktitle}{\emph{26th international conference on intelligent user interfaces}}. \bibinfo{pages}{318--328}.
\newblock


\bibitem[Weidinger et~al\mbox{.}(2021)]%
        {weidinger2021ethical}
\bibfield{author}{\bibinfo{person}{Laura Weidinger}, \bibinfo{person}{John Mellor}, \bibinfo{person}{Maribeth Rauh}, \bibinfo{person}{Conor Griffin}, \bibinfo{person}{Jonathan Uesato}, \bibinfo{person}{Po-Sen Huang}, \bibinfo{person}{Myra Cheng}, \bibinfo{person}{Mia Glaese}, \bibinfo{person}{Borja Balle}, \bibinfo{person}{Atoosa Kasirzadeh}, {et~al\mbox{.}}} \bibinfo{year}{2021}\natexlab{}.
\newblock \showarticletitle{Ethical and social risks of harm from language models}.
\newblock \bibinfo{journal}{\emph{arXiv preprint arXiv:2112.04359}} (\bibinfo{year}{2021}).
\newblock


\bibitem[Windschitl et~al\mbox{.}(2013)]%
        {windschitl2013so}
\bibfield{author}{\bibinfo{person}{Paul~D Windschitl}, \bibinfo{person}{Aaron~M Scherer}, \bibinfo{person}{Andrew~R Smith}, {and} \bibinfo{person}{Jason~P Rose}.} \bibinfo{year}{2013}\natexlab{}.
\newblock \showarticletitle{Why so confident? The influence of outcome desirability on selective exposure and likelihood judgment}.
\newblock \bibinfo{journal}{\emph{Organizational Behavior and Human Decision Processes}} \bibinfo{volume}{120}, \bibinfo{number}{1} (\bibinfo{year}{2013}), \bibinfo{pages}{73--86}.
\newblock


\bibitem[Yang et~al\mbox{.}(2019)]%
        {yang2019xfake}
\bibfield{author}{\bibinfo{person}{Fan Yang}, \bibinfo{person}{Shiva~K Pentyala}, \bibinfo{person}{Sina Mohseni}, \bibinfo{person}{Mengnan Du}, \bibinfo{person}{Hao Yuan}, \bibinfo{person}{Rhema Linder}, \bibinfo{person}{Eric~D Ragan}, \bibinfo{person}{Shuiwang Ji}, {and} \bibinfo{person}{Xia Hu}.} \bibinfo{year}{2019}\natexlab{}.
\newblock \showarticletitle{Xfake: Explainable fake news detector with visualizations}. In \bibinfo{booktitle}{\emph{The world wide web conference}}. \bibinfo{pages}{3600--3604}.
\newblock


\bibitem[Zade et~al\mbox{.}(2023)]%
        {zade2023tweet}
\bibfield{author}{\bibinfo{person}{Himanshu Zade}, \bibinfo{person}{Megan Woodruff}, \bibinfo{person}{Erika Johnson}, \bibinfo{person}{Mariah Stanley}, \bibinfo{person}{Zhennan Zhou}, \bibinfo{person}{Minh~Tu Huynh}, \bibinfo{person}{Alissa~Elizabeth Acheson}, \bibinfo{person}{Gary Hsieh}, {and} \bibinfo{person}{Kate Starbird}.} \bibinfo{year}{2023}\natexlab{}.
\newblock \showarticletitle{Tweet Trajectory and AMPS-based Contextual Cues can Help Users Identify Misinformation}.
\newblock \bibinfo{journal}{\emph{Proceedings of the ACM on Human-Computer Interaction}} \bibinfo{volume}{7}, \bibinfo{number}{CSCW1} (\bibinfo{year}{2023}), \bibinfo{pages}{1--27}.
\newblock


\bibitem[Zhang et~al\mbox{.}(2018)]%
        {zhang2018fauxbuster}
\bibfield{author}{\bibinfo{person}{Daniel~Yue Zhang}, \bibinfo{person}{Lanyu Shang}, \bibinfo{person}{Biao Geng}, \bibinfo{person}{Shuyue Lai}, \bibinfo{person}{Ke Li}, \bibinfo{person}{Hongmin Zhu}, \bibinfo{person}{Md~Tanvir Amin}, {and} \bibinfo{person}{Dong Wang}.} \bibinfo{year}{2018}\natexlab{}.
\newblock \showarticletitle{Fauxbuster: A content-free fauxtography detector using social media comments}. In \bibinfo{booktitle}{\emph{2018 IEEE international conference on big data (big data)}}. IEEE, \bibinfo{pages}{891--900}.
\newblock


\end{thebibliography}
